\date{}
\documentclass[amssymb,prd,superscriptaddress,aps,nofootinbib,twocolumn,preprintnumbers]{revtex4-1}
\usepackage[utf8]{inputenc}
\usepackage{amsmath}
\usepackage{amssymb}
\usepackage{xcolor} 
\usepackage{graphicx, epsfig, amssymb} 
\usepackage{moresize}
\usepackage[normalem]{ulem}

\usepackage{mathrsfs} 
\usepackage{tensor}
\usepackage{graphicx} \usepackage{bm} \usepackage{epsfig}

\usepackage[centerlast]{caption}
\usepackage{subcaption}
\usepackage{natbib}
\usepackage[english]{babel}
\usepackage{appendix}

\usepackage[utf8]{inputenc} 

\usepackage{graphicx}

\usepackage{subcaption}
\usepackage{float}
\usepackage{amsmath}
\usepackage{upgreek}
\usepackage{dsfont}
\usepackage{physics}

\usepackage{breakcites}
\usepackage[breaklinks=true]{hyperref}

\usepackage{hyperref}

\setcitestyle{square}

\begin{document}

\title{Quantum backreaction of $O(N)$-symmetric  scalar fields and de Sitter spacetimes at the renormalization point: Renormalization schemes and the screening of the cosmological constant}

\author{Diana L. López Nacir}
\author{Julián Rovner}
  
\affiliation{Departamento de F\'i­sica  Juan Jos\'e Giambiagi, FCEyN UBA and IFIBA CONICET-UBA,
 Facultad de Ciencias Exactas y Naturales\\Ciudad Universitaria, Pabell\'on I, 1428 Buenos Aires, Argentina}

\date{\today}

\begin{abstract}
We consider a theory of $N$ self-interacting quantum scalar fields with quartic $O(N)$-symmetric potential, with a coupling constant $\lambda$, in a generic curved spacetime. 
 We   analyze the renormalization process of the semiclassical Einstein equations at leading order in the $1/N$ expansion for  different renormailzation schemes, namely: the traditional one that sets the geometry of the spacetime to be Minkowski at the renormalization point and  other schemes (originally proposed in \cite{RevisitedI,RevisitedII} ) that set the geometry  to be that of a fixed de Sitter spacetime.
In particular, we study the quantum backreaction for fields in de Sitter spacetimes with masses much smaller than the expansion rate $H$. We find that the scheme that  
   uses the classical de Sitter background solution at the renormalization point,  stands out as the most convenient to study the quantum effects on de Sitter spacetimes. 
We obtain that the backreaction is suppressed by   $H^2/M_{pl}^2$ with no logarithmic enhancement factor of  $\ln{\lambda}$, giving  only a small screening of the classical cosmological constant due to the backreaction of such quantum fields. We point out the use of the new schemes can  also be more  convenient than the traditional one to study quantum effects in other spacetimes relevant for cosmology.

\end{abstract}


\maketitle

\section{Introduction}
 
 The motivations for studying quantum fields in de Sitter (dS) spacetime are diverse. 
 The understanding of the predictions and the robustness of the
 inflationary models usually requires to assess the importance of  quantum effects  for  light scalar fields on a  (quasi) de Sitter background geometry. For a pedagogical introduction to the main issues and the relevance of the IR behavior of quantum fields in inflationary  cosmology, see Refs.~\cite{Hu:2018nxy,Hu:2020luk}.
In the base  cosmological model, the accelerated expansion of the Universe can be described by adjusting the value of the so-called cosmological constant,  $\Lambda$, for which there is no fundamental explanation nor understanding of the inferred particular value \cite{RevModPhys.61.1,Padmanabhan:2002ji,Martin:2012bt,Planck2018}. Being as $\Lambda$ is a constant, if the classical predictions of the model are extrapolated to future times, the geometry of the Universe would approach to that of dS.
There are arguments in the literature indicating the adjusted value  is too small compared to the theoretical estimates. 
An interesting concept that aims at overcoming  this discrepancy is that of the ``screening of the cosmological constant,’’ which is  based on the expectation that large infrared quantum effects  produce an effective reduction (or screening) of the  classical value \cite{Polyakov:1982ug,Myhrvold:1983hx,Ford:1984hs,Mottola:1984ar,Antoniadis:1985pj}. dS  stands out among other possible curved geometries for its symmetries,  which in quantity equal those of the flat  Minkowski spacetime. Therefore,  it should be a good starting point for the exploration of field theories in more generic backgrounds  with nonconstant   curvatures, such as  Friedman Robertson Walker spacetimes. 
 However, there are certain characteristics of dS that hinder the development of computational methods as powerful as those     known  for Minkowski. 
One of them is the breakdown of the standard perturbation theory for light quantum fields, such as for scalar field models with (self-) interaction potentials, which are widely used in cosmology  \cite{Burgess:2010dd}. 

Several  nonperturbative approaches have been considered in the literature  to address  this problem, including
stochastic formulations based on \cite{Starobinsky:1994bd,Starobinsky:1986fx,Tsamis:2005bh}, the so-called Hartree approximation  \cite{Arai:2011dd,Arai, RevisitedI,RevisitedI}, the $1/N$ expansion \cite{Mazzitelli:1988ib,Riotto:2008mv,Serreau:2011fu},  renormalization group equations \cite{Guilleux:2016oqv,Moreau:2018lmz}, or the connection to the theory formulated on the sphere (i.e., on the Euclidean version of dS space) \cite{Hu:1985uy,Hu:1986cv,Rajaraman:2010xd,Beneke:2012kn,Nacir:2016fzi,LopezNacir:2016gfj,LopezNacir:2018xto,LopezNacir:2019ord}.

In this paper we assess the impact of  choosing the ultraviolet (UV) renormalization scheme  in the physical understanding and robustness of the nonperturbative results. The approach we use is based on the work done previously in \cite{RevisitedI,RevisitedII}.  
This consists in using  the two-particle irreducible  effective action   (2PI-EA) method \cite{CalzettaHu}  to   derive finite (renormalized) self-consistent equations for  the mean fields, the two point functions, and the metric tensor $g_{\mu\nu}$, which are  nonperturbative in the (self-)coupling of the scalar fields. In Refs. \cite{RevisitedI,RevisitedII} the study was done for the more difficult case of only one scalar field, where there is no parameter controlling the truncation of the diagrammatic expansion.  
Here we focus on   the large $N$ limit of an $O(N)$-symmetric  scalar field theory.   

In Sec.~\ref{SectIntro} we summarize the nonperturbative formalism we consider, following \cite{RamseyHu}, and we  derive the effective action in the large $N$ limit. In Sec.~\ref{Sectrenorm} we critically study the renormalization process for the so-called gap equation, from which a nonperturbative dynamical mass is obtained. We consider three different renormalization schemes, namely: the minimal subtraction (MS) scheme, the Minkowski renormalization (MR) scheme and the de Sitter renormalization (dSR) scheme.  A sketch of the derivation of the renormalized semiclassical Einstein Equations (SEE) is provided in Sec.~ \ref{SecRenSEE}, which closely follows previous studies for $N=1$ field in the Hartree approximation \cite{RevisitedII} (see also \cite{RevisitedI}).
The renormalized SEE for a generic metric $g_{\mu\nu}$ are obtained in the dSR scheme, using a fixed dS metric with curvature a $R_0$. These equations reduce to the traditional renormalized SEE when $R_0\to 0$. In Sec.~\ref{SecRenSEEdS} we specialize the results for a  dS background metric, for which dS self-consistent solutions are studied in \ref{SecSelfC}.
In this study we show that, given the  role of dS curvature in generating a nonperturbative dynamical mass, the introduction of the dSR scheme is crucial in understanding the infrared effects for light quantum fields.

Our results agree with those found in Ref.~\cite{Moreau:2018lmz}, using  nonperturbative renormalization group techniques, on that the expected large  infrared corrections to the curvature of the spacetime show up as a manifestation of the breakdown of perturbation theory (rather than an instability). The corrections are screened  when nonperturbative (mainly,  the dynamical generation of a mass) effects are accounted for. Indeed, the infrared corrections to the dS curvature turn out to be  controlled by the ratio $H^2/M_{pl}^2$, which is small by assumption in the semiclassical regime. In particular, in contradiction with the results reported in \cite{Arai}, we find  no logarithmic enhancement factor of $\ln\lambda$  in the renormalized stress energy tensor  (we clarify  the reason of this disagreement in Sec. \ref{ComparacionConAraiYSerreau}).

Everywhere we set $c=\hbar=1$ and adopt the mostly plus sign convention. 

\section{The 2PI effective action at leading order in $1/N$ }\label{SectIntro}

We consider an  $O(N)$-symmetric  scalar field theory in de Sitter spacetime with an action,

\begin{equation}
\begin{aligned}\label{ModelSF}
    S^F[\phi^i,g^{\mu\nu}] = &-\frac{1}{2}\int d^dx\sqrt{-g}\left[ \phi^i(-\Box+m^2_B+\xi_B R)\phi^j\delta_{ij}\right.\\
    &\left.+\frac{\lambda_B}{4N}(\phi^i\phi^j\delta_{ij})^2 \right],
    \end{aligned}
\end{equation}
where  $i,j$ are the index of the $N$ scalar fields $\phi^i$ of the theory (where the sum convention is used), $m_B$ is the bare mass of the fields, $\xi_B$ is the bare coupling to the curvature $R$, $\lambda_B$ is the bare parameter controlling the fields (self-)interaction, and $\delta_{ij}$ is the identity matrix in $N$ dimensions.

A systematic $1/N$ expansion can be obtained in the framework of the so-called two-particle irreducible  effective action    (2PI EA). 
The definition of the 2PI EA along with the corresponding functional integral can be found in both  papers and textbooks (see, for instance,  \cite{CJT,CalzettaHu,RamseyHu,Hu:2020luk}). In this section 
we briefly  summarize the main relevant aspects of the formalism for the model we are considering and the results obtained in \cite{RamseyHu} using the $1/N$ expansion.  

We work in the “closed-time-path” (CTP) formalism \cite{CalzettaHu} and introduce a set of indexes $a,b$ that can be either $+$ or $-$ depending on the time branch.
The starting point to obtain the 2PI EA  is the introduction of  a local source $J(x)$ as well as a nonlocal one $K(x,x')$ in the generating functional $Z[J,K]$. The 2PI EA is  the double Legendre transform with respect to that sources and  is a functional of the mean fields $\hat{\phi}^i_a$ and the propagators $G^{ij}_{ab}$. The result can be written as \cite{RamseyHu}

\begin{equation}
    \begin{aligned}
    &\Gamma_{2PI}[\hat{\phi}^i,G,g^{\mu\nu}]=S^F[\hat{\phi}^i,g^{\mu\nu}]-\frac{i}{2}\ln\det[G_{ab}^{ij}]\\
     &+\frac{i}{2}\int d^4x\sqrt{-g}\int d^4x'\sqrt{-g'}\mathcal{A}_{ij}^{ab}(x',x)G_{ab}^{ij}(x,x')\\
     &+\Gamma_2[\hat{\phi^i},G,g^{\mu\nu}],
    \end{aligned}
    \label{2PIO(N)}
\end{equation}\\
where 
$\Gamma_2[\hat{\phi}^i,G,g^{\mu\nu}]$ is $-i$ times the 2PI vacuum Feynman diagrams with the propagator $\mathcal{A}^{ab}$ given by

\begin{equation}
    \left.i\mathcal{A}_{ij}^{ab}(x,x')=\frac{1}{\sqrt{-g}}\left(\frac{\delta^2S^F}{\delta\phi_a^i(x)\delta\phi_b^j(x')}\right)\frac{1}{\sqrt{-g'}}\right|_{\phi=\hat{\phi}}.
\end{equation}\\and with a vertex defined by $\mathcal{S}_{int}^F[\varphi^i,g^{\mu\nu}]$, which is the  interaction action obtained after recollecting the cubic and higher orders in $\hat{\phi}^i$ that emerged when  expanding ${S}^F[\hat{\phi}^i+\varphi^i,g^{\mu\nu}]$ (with $\varphi^i=\phi^i-\hat{\phi}^i)$,

\begin{equation}
\begin{aligned}
    \mathcal{S}^F_{int}[\varphi^i,g^{\mu\nu}]&=-\frac{\lambda_B}{2N}\int d^dx\sqrt{-g}\left[ \frac{1}{4}(\varphi^i\varphi^j\delta_{ij})^2\right.\\
    &\left.+(\hat{\phi}^i\varphi^j\delta_{ij}).(\hat{\phi}^k\varphi^l\delta_{kl}) \right],
    \end{aligned}
\end{equation}

Following \cite{RamseyHu}, the leading order in   the $1/N$ expansion in the unbroken symmetry case ($\langle \phi^i\rangle=0$) result:

  \begin{equation}
    \begin{aligned}
    &\Gamma_{2PI}[\hat{\phi},G,g^{\mu\nu}]=\mathcal{S}^F[\hat{\phi},g^{\mu\nu}]-\frac{iN}{2}\ln\det[G_{ab}]\\
     &+\frac{iN}{2}\int d^4x\sqrt{-g}_a\int d^4x'\sqrt{-g}_{b}\mathcal{A}^{ab}(x',x)G_{ab}(x,x')\\
      &-\frac{\lambda_B N}{8}c^{abcde}\int d^4x\sqrt{-g}_eG_{ab}(x,x)G_{cd}(x,x)+\mathcal{O}(1),
    \end{aligned}
    \label{2PIExpansion1/N}
\end{equation}where $c^{abcde}$ is equal to $\pm 1$, if $a=b=c=d=e=\pm$, and zero otherwise.

By setting to zero the variation of this 2PI EA with respect to the scalar field and the propagators, we obtain
\begin{equation}
    \left( -\Box+m_B^2+\xi_B R+\frac{\lambda_B}{2}\hat{\phi}^2+\frac{\lambda_B}{4}[G_1] \right)\hat{\phi}(x)=0,
    \label{EquMovPhi}
\end{equation}
\begin{equation}
    \left( -\Box+m_B^2+\xi_B R+\frac{\lambda_B}{2}{\hat{\phi}}^2+\frac{\lambda_B}{4}[G_1] \right)G_{1}(x,x')=0,
    \label{EquMovG}
\end{equation} where $\hat{\phi}^2=\hat{\phi}_i\hat{\phi}_j\delta_{ij}/N$ and $[G_1]=2G(x,x)=2G_{ij}(x,x)\delta_{ij}/N$ is the coincidence limit of the Hadamard propagator, which is a divergent quantity.  Therefore, the two equations above contain divergences. In this paper we  use dimensional regularization together with an adiabatic expansion. In the next section we analyze the  renormalization process.

     \section{Renormalization in curved spacetimes and renormalization schemes}\label{Sectrenorm}

  In this section we describe the renormalization process of Eqs.~(\ref{EquMovPhi}) and (\ref{EquMovG}) in three different renormalization schemes, which we refer to as: the minimal subtraction (MS) scheme, the Minkowski renormalization (MR) scheme and the de Sitter renormalization (dSR) schemes. 
   
   \subsection{Minimal subtraction scheme}

  In  the  MS scheme, we split each   bare parameter  ($m^2_B$, $\xi_B$, $\lambda_B$) into  the MS scheme  parameter ($m^2$, $\xi$, $\lambda$), which defines the finite part, and the divergent contribution  ($\delta m^2$, $\delta\xi$, $\delta\lambda$), which contains only divergences and no finite part, 
  
  \begin{equation}
    \begin{aligned}
    m_B^2&=m^2+\delta m^2, \\
    \xi_B& = \xi+\delta\xi, \\
    \lambda_B&=\lambda+\delta\lambda .
    \end{aligned}
    \label{ParametrosDesnudos}
\end{equation}
In order to renormalize the equations we rewrite them  as,
  \begin{equation}
    \begin{aligned}
    (-\Box+m_{ph}^2+\xi_R R)&\hat{\phi}(x)=0,\\
    (-\Box+m_{ph}^2+\xi_R R)&G_1(x,x')=0,
    \end{aligned}
    \label{EqsMovConMphys}
\end{equation} defining  the  physical mass $m_{ph}$ as the solution to the following gap equation:
  \begin{equation}\label{Defmph}
  \begin{aligned}
    m_{ph}^2+\xi_R R&=(m^2+\delta m^2)+(\xi+\delta\xi)R+\frac{1}{2}(\lambda+\delta\lambda)\hat{\phi}^2\\
    &+\frac{1}{4}(\lambda+\delta\lambda)[G_1].
    \end{aligned}
\end{equation}
We  now use the well-known Schwinger-DeWitt expansion for $[G_1]$ to split the propagator into its divergent and finite terms \cite{Mazzitelli:1988ib},

\begin{equation}\label{G1coincidence}
    [G_1]=\frac{1}{4\pi^2\epsilon}\left[m_{ph}^2+\left(\xi_R-\frac{1}{6}\right)R\right]+2T_F(m_{ph}^2,\xi_R,\{R\},\hat{\mu}),
\end{equation}
where $\epsilon=d-4$, which is factored out in the first term making explicit the divergence as $d\to4$, $T_F$ is a finite function that depends on the spacetime (here $\{R\}$ denotes the curvature tensors), where a scale $\hat{\mu}$ with units of mass must be included in order for the physical quantities to have the correct units along the dimensional regularization procedure.

Introducing Eq.~(\ref{G1coincidence}) into Eq.~(\ref{Defmph}) and   demanding that the divergent terms cancel  out with the contributions of the counterterms, we obtain a finite equation for the  physical mass (i.e., the renormalized gap equation):
  \begin{equation}
    \begin{aligned}
    &m_{ph}^2=m^2+\frac{1}{2}\lambda\hat{\phi}^2\\
    &+\frac{\lambda}{32\pi^2}\left\{ \left[ m_{ph}^2+\left( \xi_R-\frac{1}{6} \right)R \right]\ln\left( \frac{m_{ph}^2}{\hat{\mu}^2} \right)\right.\\
     &+\left.\left( \xi_R-\frac{1}{6}
    \right)R-2F(m_{ph}^2,\{R\})\right\} ,
    \end{aligned}
    \label{mphSinParamRenorm}
\end{equation} where the function $F(m_{ph}^2,\{R\})$   is defined by

\begin{equation}
    \begin{aligned}
    T_F(m^2,\xi_R,\{R\},\hat{\mu})&=\frac{1}{16\pi^2}\left\{\left[ m^2+\left(\xi-\frac{1}{6}\right)R \right]\ln\left(\frac{m^2}{\hat{\mu}^2}\right)  \right.\\
    &\left. \left(\xi-\frac{1}{6}\right)R-2F(m,\{R\}) \right\},
    \end{aligned}
    \label{TF}
\end{equation} and  has the following properties:

\begin{equation}
    \begin{aligned}
    \left.F(m^2,\{R\})\right|_{R_{\mu\nu\rho\sigma}=0}&=0,\\
    \left.\frac{dF(m^2,\{R\})}{dm^2}\right|_{R_{\mu\nu\rho\sigma}=0}&=0,\\
    \left.\frac{dF(m^2,\{R\})}{dR}\right|_{R_{\mu\nu\rho\sigma}=0}&=0.
    \end{aligned}
\end{equation}

In Appendix A we provide the expression for the counterterms.

Therefore, the gap equation  in the MS scheme depends on the mass scale $\hat{\mu}$, which is an  arbitrary scale with no obvious physical interpretation. A way to  define  renormalized parameters with a physical meaning  is to use the effective potential\footnote{Alternatively, one can use the stress-energy tensor; see, for instance \cite{Markkanen:2013nwa}.  }, which can be   obtained by integrating the following equation with respect to $\hat{\phi}$, following Eq. \eqref{EqsMovConMphys}: 

  \begin{equation}
    \frac{dV_{eff}}{d\hat{\phi}}=(m_{ph}^2+\xi_R R)\hat{\phi}.
    \label{PotencialEfectivo}
\end{equation}  In this way,   a natural definition for the renormalized parameters ($m_R$, $\xi_R$  $\lambda_R$)
is to set them  to be equal to the corresponding derivatives of the effective potential as a function of $\hat{\phi}$ and $R$ evaluated at $\hat{\phi}=0$ and $R=0$ (that is, as defined in Minkowski space) and more generally at $R=R_0$. This is the option we adopt next.

 \subsection{Minkowski renormalization  scheme }
 
 Choosing Minkowski geometry at  the renormalization point, corresponds  to  setting $R_0$ to zero. Therefore, from the effective potential $V_{eff}$ the renormalized parameters are obtained from Eq. \eqref{mphSinParamRenorm} as follows:
 
 \begin{equation}
    m_R^2\equiv\left. \frac{d^2V_{eff}}{d\hat{\phi}^2} \right|_{\hat{\phi}=0,R=0}=m_{ph}^2\Bigg{|}_{\hat{\phi}=0,R=0},
\end{equation}

\begin{equation}
    \xi_R\equiv\left. \frac{d^3V_{eff}}{dRd\hat{\phi}^2} \right|_{R=0}=\left.\frac{dm_{ph}^2}{dR}\right|_{R=0}+\xi_R,
\end{equation}

\begin{equation}
    \lambda_R\equiv\left. \frac{d^4V_{eff}}{d\hat{\phi}^4} \right|_{\hat{\phi}=0,R=0}=3\left. \frac{d^2m_{ph}^2}{d\hat{\phi}^2} \right|_{\hat{\phi}=0,R=0}.
\end{equation}

And the result is:

\begin{equation}
    m_R^2=\frac{m^2}{\left[ 1-\frac{\lambda}{32\pi^2}\ln\left( \frac{m_R^2}{\hat{\mu}^2} \right) \right]},
    \label{mR2R00}
\end{equation}

\begin{equation}
    \left( \xi_R-\frac{1}{6} \right)=\frac{\left( \xi-\frac{1}{6} \right)}{\left[ 1-\frac{\lambda}{32\pi^2}-\frac{\lambda}{32\pi^2}\ln\left( \frac{m_R^2}{\hat{\mu}^2} \right) \right]},
    \label{xiRR00}
\end{equation}

\begin{equation}
    \lambda_R=\frac{3\lambda}{\left[ 1-\frac{\lambda}{32\pi^2}-\frac{\lambda}{32\pi^2}\ln\left( \frac{m_R^2}{\hat{\mu}^2} \right) \right]}.
    \label{lambdaRR00}
\end{equation}

From these equations, we can find the following useful relations between   the MR parameters defined above and the MS parameters,

\begin{equation}
    \frac{\left(\xi_B-\frac{1}{6}\right)}{\lambda_B}=\frac{\left( \xi-\frac{1}{6} \right)}{\lambda}=\frac{3\left( \xi_R-\frac{1}{6} \right)}{\lambda_R},
\end{equation}

\begin{equation}
    \frac{m_B^2}{\lambda_B}=\frac{m^2}{\lambda}=m_R^2\left( \frac{1}{32\pi^2}+\frac{3}{\lambda_R} \right)\equiv \frac{m_R^2}{\lambda_R^*},
\end{equation}
\noindent where we introduced $\lambda_R^*$ to simplify the   notation. Then, we can write the equation for $m_{ph}^2$, using solely one set of parameters,

\begin{equation}
\begin{aligned}
    m_{ph}^2=m_R^2&+\frac{\lambda_R^*}{2}\hat{\phi}^2\\
    &+\frac{\lambda_R^*}{32\pi^2}\left\{ \left[ m_{ph}^2+\left( \xi_R-\frac{1}{6} \right)R \right]\ln\left( \frac{m_{ph}^2}{m_R^2} \right)\right.\\
    &-2F(m_{ph}^2,\{R\}) \Bigg{\}}.
    \label{MphR00}
    \end{aligned}
\end{equation}

One can easily check that, in the free field limit ($\lambda_R\hspace{0.1cm}\rightarrow\hspace{0.1cm}0$, and therefore $\lambda_R^*\hspace{0.1cm}\rightarrow\hspace{0.1cm}0$)  the physical mass  reduces to the renormalized mass, $m_{ph}^2\hspace{0.1cm}\rightarrow\hspace{0.1cm}m_R^2$.
 
\subsection{de Sitter renormalization schemes }\label{dSR}
 
 Now we set the geometry of the spacetime at the renormalization point to be that of a fixed de Sitter spacetime, corresponding to $R=R_0=$ constant. As above,  the   renormalized parameters ($m_R^2$, $\xi_R$, $\lambda_R$) are defined in terms of the  effective potential,
\begin{equation}
    m_R^2=\left. \frac{d^2V_{eff}}{d\hat{\phi}^2} \right|_{\hat{\phi}=0, R=R_0}=\left. m_{ph}^2 \right|_{\hat{\phi}=0, R=R_0}+\xi_R R_0,
\end{equation}

\begin{equation}
    \xi_R=\left. \frac{d^3V_{eff}}{dRd\hat{\phi}^2} \right|_{\hat{\phi}=0, R=R_0}=\left. \frac{dm_{ph}^2}{dR} \right|_{\hat{\phi}=0, R=R_0}+\xi_R,
\end{equation}

\begin{equation}
    \lambda_R=\left. \frac{d^4V_{eff}}{d\hat{\phi}^4} \right|_{\hat{\phi}=0, R=R_0}=3\left.\frac{dm_{ph}^2}{d\hat{\phi}^2}  \right|_{\hat{\phi}=0, R=R_0}.
\end{equation}\\
Therefore, the generalization of the expressions (\ref{mR2R00}), (\ref{xiRR00}) and (\ref{lambdaRR00})  that relate these parameters to the MS parameters are:

\begin{equation}
\begin{aligned}
   & \left[ 1-\frac{\lambda}{32\pi^2}\ln\left(\frac{m_R^2}{\hat{\mu}^2} \right) \right]m_R^2=(\xi-\xi_R)R_0+m^2\\
   &+\frac{\lambda}{32\pi^2}\left[   (\xi_R-1/6)R_0\left( 1+\ln\left( \frac{m_R^2}{\hat{\mu}^2} \right) \right)\right.\\
   &-2F_{dS}(m_R^2,R_0)\Bigg{]},
    \label{mR2R0cte}
    \end{aligned}
\end{equation}

\begin{equation}
    \left( \xi_R-\frac{1}{6} \right)=\frac{\left( \xi-\frac{1}{6} \right)-\frac{\lambda}{16\pi^2}\left. \frac{dF_{dS}}{dR} \right|_{m_R^2,R=R_0}}{\left[ 1-\frac{\lambda}{32\pi^2}-\frac{\lambda}{32\pi^2}\ln\left( \frac{m_{R}^2}{\hat{\mu}^2} \right) \right]},
    \label{xiRR0cte}
\end{equation}

\begin{equation}
\begin{aligned}
    &\lambda_R=3\lambda\left[1-\frac{\lambda}{32\pi^2}\left\{1+\ln\left( \frac{m_{R}^2}{\hat{\mu}^2}\right)\right.\right.\\
    &\left.\left.+\frac{(\xi_R-1/6)R_0}{m_{R}^2}-\left. 2\frac{dF_{dS}}{dm^2} \right|_{m_R^2,R=R_0}  \right\} \right]^{-1}
    \label{lambdaRR0cte}.
    \end{aligned}
\end{equation}

As for the MR case,  one can derive the following relations between these renormalized parameters and the MS  ones:
\begin{equation}
\begin{aligned}
    \frac{m_B^2}{\lambda_B}=\frac{m^2}{\lambda}&=m_R^2\left( \frac{1}{32\pi^2}+\frac{3}{\lambda_R} \right)+\frac{(\xi_R-1/6)R_0}{32\pi^2}\\
    &\equiv\frac{m_R^2}{\lambda_R^*}+\frac{(\xi_R-1/6)R_0}{32\pi^2},
    \label{lambdaRestrella}
    \end{aligned}
\end{equation}

\begin{equation}
    \begin{aligned}
    &\frac{(\xi_B-1/6)}{\lambda_B}=\frac{(\xi-1/6)}{\lambda}=3\frac{(\xi_R-1/6)}{\lambda_R}\\
    &+\frac{1}{16\pi^2}\left. \frac{dF_{dS}}{dR}\right|_{m_R^2,R=R_0}\\
     &+\frac{(\xi_R-1/6)}{32\pi^2}\left[ \frac{(\xi_R-1/6)R_0}{m_R^2}-2\left. \frac{dF_{dS}}{dm^2}\right|_{m_R^2,R=R_0} \right]\\
      &\equiv3\frac{(\xi_R-1/6)}{\lambda_R}+J(R_0,m_R^2,\xi_R),
    \end{aligned}
    \label{RelacionUtilxiRlambdaR}
\end{equation}

\noindent where the function $J(R_0,m_R^2,\xi_R)$ is defined by the last equality and goes to zero when $R_0\hspace{0.1cm}\rightarrow\hspace{0.1cm}0$.

Then, using the Eqs. (\ref{mR2R0cte}), (\ref{xiRR0cte}) and (\ref{lambdaRR0cte}), the new expression for the physical mass $m_{ph}^2$ is found,

\begin{equation}
    \begin{aligned}
    &m_{ph}^2=m_R^2+\frac{\lambda_R^*}{32\pi^2}\left\{\left[m_{ph}^2+\left( \xi_R-\frac{1}{6} \right)R \right]\ln\left( \frac{m_{ph}^2}{m_R^2} \right)\right.\\
     &+(m_{ph}^2-m_R^2)\left[ 2\left.\frac{dF_{dS}}{dm_{ph}^2}\right|_{m_R^2,R_0}-\frac{(\xi_R-1/6)R_0}{m_R^2} \right]\\
      &+2\left[ F_{dS}(m_R^2,R_0)+\left.\frac{dF_{dS}}{dR}\right|_{m_R^2,R_0}(R-R_0)\right.\\
      &-F(m_{ph}^2,\{R\}) \Bigg{]}\Bigg{\}}+\frac{\lambda_R^*}{2}\hat{\phi}^2.
    \end{aligned}
    \label{mphR0cte}
\end{equation}

This result reduces to the previous one in the MR scheme, given in (\ref{MphR00}), when $R_0\to0$. Finally, the resulting counterterms are given by

\begin{equation}
\begin{aligned}
    &\delta \overline{m}^2\equiv m_B^2-m_R^2\\
    &=-\frac{m_B^2m_R^2}{32\pi^2}\frac{\left[ \frac{2}{\epsilon}+\ln\left( \frac{m_R^2}{\hat{\mu}^2} \right)-2\left.\frac{dF_{dS}}{dm^2}\right|_{m_R^2,R_0} \right]}{\left( \frac{m_R^2}{\lambda_R^*}+\frac{(\xi_R-1/6)R_0}{32\pi^2} \right)},
    \label{deltamR0cte}
    \end{aligned}
\end{equation}

\begin{equation}\begin{aligned}
    &\delta \overline{\xi}\equiv\xi_B-\xi_R=-\frac{(\xi_B-\frac{1}{6})}{32\pi^2}\times\\
    &\frac{\left\{ (\xi_R-\frac{1}{6})\left[ \frac{2}{\epsilon}+1+\ln\left( \frac{m_R^2}{\hat{\mu}^2} \right) \right] +2\left. \frac{dF_{dS}}{dR} \right|_{m_R^2,R_0} \right\}}{\left[ \frac{3(\xi_R-\frac{1}{6})}{\lambda_R}+J \right]},
    \label{deltaxiR0cte}
    \end{aligned}
\end{equation}

\begin{equation}
\begin{aligned}
    \delta\overline{\lambda}\equiv&(\lambda_B-\lambda_R)\\
    &=-2\lambda_B-\frac{\lambda_B\lambda_R}{32\pi^2}\left[ \frac{2}{\epsilon}+1+\ln\left( \frac{m_R^2}{\hat{\mu}^2} \right)\right.\\
    &\left.+\frac{(\xi_R-\frac{1}{6})R_0}{m_R^2}-2\left.\frac{dF_{dS}}{dm^2}\right|_{m_R^2,R_0} \right]
    \label{deltalambdaR0cte}.
    \end{aligned}
\end{equation}

Before proceeding any further, we present the expression of the  function $F_{dS}(m^2,R)$ which is the one defined in \eqref{TF}  evaluated in the de Sitter spacetime. To see a more detailed derivation we encourage the reader to read \cite{RevisitedI,RevisitedII},

\begin{equation}
    \begin{aligned}
    &F_{dS}(m^2,R)=-\frac{R}{2}\left\{ \left(\frac{m^2}{R}+\xi-\frac{1}{6}\right)\times\right.\\
    &\left[\ln\left(\frac{R}{12m^2}\right)+g(m^2/R+\xi)\right]
      \left. -\left(\xi-\frac{1}{6}\right)-\frac{1}{18} \right\},
    \end{aligned}
    \label{FdS}
\end{equation}

with

\begin{equation}
    g\left(\frac{m^2+\xi R}{R}\right)\equiv \psi_++\psi_-=\psi\left(\frac{3}{2}+\nu_4\right)+\psi\left(\frac{3}{2}-\nu_4\right),
    \label{g}
\end{equation}
where $\psi(x)=\Gamma'(x)/\Gamma(x)$ is the DiGamma function and we define $\nu_4\equiv\sqrt{9/4-12(m^2+\xi R)/R}$. One important property of the function $g$ is that, in the infrared limit $m^2+\xi R\ll R$,

\begin{equation}
    \begin{aligned}
    g\left(\frac{m^2+\xi R}{R}\right)&\simeq-\frac{R}{4(m^2+\xi R)}+\frac{11}{6}\\
     &-2\gamma_E+\frac{49}{9}\frac{(m^2+\xi R)}{R}.
    \end{aligned}
    \label{gLimIR}
\end{equation}

\section{Renormalization of the semiclassical Einstein equations}\label{SecRenSEE}
We are halfway to our goal; the procedure below corresponds to the other half. The   equations  obtained above from the 2PI EA, describe the dynamics of $\hat{\phi}$ and $G$ for a generic metric $g_{\mu\nu}$. In order to assess the effect of the quantum fields on the spacetime geometry, we need to set to zero  the variation of the  2PI EA action  including the gravitational part with respect to $g_{\mu\nu}$. This is equivalent to computing
the expectation value of the stress-energy tensor $\langle T_{\mu\nu}\rangle$ and  use it as a source in the semiclassical Einstein equations (SEE). The resulting SEE are given by \cite{Mazzitelli:1988ib}
\begin{equation}
\begin{aligned}
    &k_B^{-1}G_{\mu\nu}+\Lambda_Bk_{B}^{-1}g_{\mu\nu}+\alpha_{1B}^{(1)}H_{\mu\nu}+\alpha_{2B}^{(2)}H_{\mu\nu}+\alpha_{3B}H_{\mu\nu}\\
    &=\langle T_{\mu\nu}\rangle,
    \end{aligned}
\end{equation}

where

\begin{equation}
    ^{(1)}H_{\mu\nu}=2R_{;\mu\nu}-2g_{\mu\nu}\Box R+\frac{1}{2}g_{\mu\nu}R^2-2RR_{\mu\nu},
\end{equation}

\begin{equation}
\begin{aligned}
    &^{(2)}H_{\mu\nu}=R_{;\mu\nu}-\frac{1}{2}g_{\mu\nu}\Box R-\Box R_{\mu\nu}+\frac{1}{2}g_{\mu\nu}R^{\alpha\beta}R_{\alpha\beta}\\
    &-2R^{\alpha\beta}R_{\alpha\beta\mu\nu},
    \end{aligned}
\end{equation}

\begin{equation}
    \begin{aligned}
    H_{\mu\nu}&=\frac{1}{2}g_{\mu\nu}R^{\rho\sigma\gamma\delta}R_{\rho\sigma\gamma\delta}-2R_{\mu\rho\sigma\gamma}R\indices{_{\nu}^{\rho\sigma\gamma}}-4\Box R_{\mu\nu}\\
     &+2R_{\mu\nu}+4R_{\mu\rho}R\indices{^{\rho}_{\nu}}+4R^{\rho\sigma}R_{\rho\mu\sigma\nu}.
    \end{aligned}
    \label{Hmunu}
\end{equation}

When the dimension is set to $d=4$,  the Gauss-Bonet theorem  implies that these tensors are not all independent from each other, and hence we have that \cite{Mazzitelli:1988ib} $H_{\mu\nu}=-^{(1)}H_{\mu\nu}+4^{(2)}H_{\mu\nu}$. 

The stress energy tensor $\langle T_{\mu\nu}\rangle$ in the large $N$ approximation, can be obtained   from the 2PI EA in Eq. (\ref{2PIExpansion1/N}). The   computation  is described in \cite{RamseyHu} and \cite{RevisitedII}, and the result for a generic metric is

\begin{equation}
    \langle T_{\mu\nu} \rangle=N\langle {\Tilde T_{\mu\nu}}\rangle_B+\frac{\lambda_B N}{32}[G_1]^2,
    \label{ValorExpectTmunu}
\end{equation}

\noindent where

\begin{equation}\label{ValorExpectTmunu2}
\begin{aligned}
    &\langle {\Tilde T_{\mu\nu}}\rangle_B = -\frac{1}{2}[G_{1;\mu\nu}]+\left( \frac{1}{4}-\frac{\xi_B}{2}\right)[G_1]_{;\mu\nu}\\
    &+\left( \xi_B-\frac{1}{4} \right)\frac{g_{\mu\nu}}{2}\Box[G_1]+\frac{1}{2}\xi_B R_{\mu\nu}[G_1],
    \end{aligned}
\end{equation}

\noindent with square brackets denoting the coincidence limit (see, for instance, \cite{PhysRevD.14.2490} for the formal definition of such limit) and the index $B$ only states that the parameters there involved are the bare ones. 

Let us now separate $\langle {\Tilde T_{\mu\nu}}\rangle_B$ into

\begin{equation}
    \langle {\Tilde T_{\mu\nu}}\rangle_B=\langle{\Tilde  T_{\mu\nu}}\rangle_R+\frac{\delta \overline{\xi}}{2}(-[G_1]_{;\mu\nu}+g_{\mu\nu}\Box [G_1]+R_{\mu\nu}[G_1])
\end{equation} where $\langle {\Tilde T_{\mu\nu}}\rangle_R$ depends not only on  the renormalized parameters,   but contains divergences coming from $G_1$ and its derivatives. As  is well known, these divergent contributions can be properly isolated by   computing the adiabatic expansion of $\langle {\Tilde T_{\mu\nu}}\rangle_R$ up to the fourth order.  The sum of such contributions are a tensor we call $\langle{\Tilde {T}_{\mu\nu}}\rangle_{ad4}$, which is given by~\cite{Mazzitelli:1988ib,RamseyHu}

\begin{equation}
    \begin{aligned}
    &\langle {\Tilde{T}_{\mu\nu}}\rangle_{ad4}=\frac{1}{16\pi^2}\left( \frac{m_{ph}^2}{\mu^2} \right)^{\epsilon/2}\left[ \frac{1}{2}\Gamma\left( -2-\frac{\epsilon}{2}\right)m_{ph}^4g_{\mu\nu}\right.\\
     &+\left. m_{ph}^2\Gamma\left( -1-\frac{\epsilon}{2} \right)\left\{ \frac{1}{2}[\Omega_1]g_{\mu\nu}+\left( \xi_R-\frac{1}{6} \right)R_{\mu\nu} \right\} \right.\\
      &+\left. \Gamma\left( -\frac{\epsilon}{2} \right) \left\{ \left( \xi_R-\frac{1}{6}\right)R_{\mu\nu}[\Omega_1]-[\Omega_{1;\mu\nu}]\right.\right.\\
      &\left.\left.\left.+\left(\frac{1}{2}-\xi_R  \right)[\Omega_1]_{;\mu\nu}\right. \right.+\left( \xi_R-\frac{1}{4} \right)g_{\mu\nu}\Box[\Omega_1] \right\}\\
      &\left.+\frac{1}{2}\Gamma\left(-\frac{\epsilon}{2} \right)[\Omega_2]g_{\mu\nu}  \right]\\
    \end{aligned}
    \label{Tmunuad4}
\end{equation} 

\noindent where the expressions for $[\Omega_1]$, $[\Omega_2]$, and $[\Omega_{1;\mu\nu}]$ are 

\begin{equation}
    [\Omega_1]=\left( \frac{1}{6}-\xi_R \right)R,
\end{equation}

\begin{equation}
\begin{aligned}
    &[\Omega_2]=\frac{1}{180}(R^{\alpha\beta\mu\nu}R_{\alpha\beta\mu\nu}-R_{\mu\nu}R^{\mu\nu})\\
    &+\frac{1}{2}R^2\left( \frac{1}{6}-\xi_R \right)^2+\frac{1}{6}\left( \frac{1}{5}-\xi_R \right)\Box R,
    \end{aligned}
\end{equation}

\begin{equation}
\begin{aligned}
    &[\Omega_{1;\mu\nu}]=\frac{1}{3}\left( \frac{3}{20}-\xi_R \right)R_{;\mu\nu}+\frac{1}{60}\Box R_{\mu\nu}\\
    &-\frac{1}{45}R_{\mu\alpha}R\indices{^{\alpha}_{\nu}}+\frac{1}{90}\left( R_{\mu\alpha\nu\beta}R^{\alpha\beta}+R_{\mu\alpha\beta\gamma}R\indices{_{\nu}^{\alpha\beta\gamma}}  \right).
    \end{aligned}
\end{equation}

The renormalization process follows closely that described in Ref.~\cite{RevisitedII} for $N=1$ in the Hartree approximation. To proceed we need to use the counterterms for  ($m_B$, $\xi_B$,  $\lambda_B$) obtained as  described above. In what follows we write the results in terms of the renormalized parameters ($m_R$, $\xi_R$,  $\lambda_R$) in the dSR scheme. The expressions in the other schemes can be found using the relations derived in the previous section. Then, by separating the full expression of the fourth adiabatic order of the $\langle {{T}_{\mu\nu}}\rangle$ given in Eq. (\ref{ValorExpectTmunu})  (which we call $\langle {{T}_{\mu\nu}}\rangle_{ad4}$) into its divergent and finite terms, we can write  $\langle {{T}_{\mu\nu}}\rangle_{ad4}=\langle{{T}_{\mu\nu}}\rangle_{ad4}^{div}+\langle{{T}_{\mu\nu}}\rangle_{ad4}^{con}$. After performing carefully the limit when $\epsilon\to0$ of $\Gamma(-\epsilon/2-a)x^{\epsilon/2}$, the convergent part results \cite{RevisitedII}
\begin{equation}
    \begin{aligned}
    &\langle {{T}_{\mu\nu}} \rangle_{ad4}^{con}=N\left\{\left(\frac{m_R^2}{2}-m_{ph}^2 \right)\left[ \frac{m_R^2}{\lambda_R^*}+\frac{(\xi_R-\frac{1}{6})R_0}{32\pi^2} \right]g_{\mu\nu} \right.\\
    &+\frac{m_{ph}^4}{64\pi^2}\left[ \frac{32\pi^2}{\lambda_R^*}+\frac{1}{2}+\left( \xi_R-\frac{1}{6} \right)\frac{R_0}{m_R^2} -2\left.\frac{dF_{dS}}{dm_{ph}^2}\right|_{m_R^2,R_0} \right]g_{\mu\nu}\\
    &+\frac{1}{16\pi^2}\left[ 2m_{ph}^2G_{\mu\nu}-\left(\xi_R-\frac{1}{6}\right) ^{(1)}H_{\mu\nu}\right]\left.\frac{dF_{dS}}{dR} \right|_{m_R^2,R_0}\\
    &+\frac{1}{32\pi^2}\ln\left(\frac{m_{ph}^2}{m_R^2}\right)\left[ -\frac{m_{ph}^4}{2}g_{\mu\nu}+2m_{ph}^2\left(\xi_R-\frac{1}{6}\right)G_{\mu\nu}\right.\\
    &\left.+\frac{1}{90}(^{(2)}H_{\mu\nu}-H_{\mu})-^{(1)}H_{\mu\nu}\left(\xi_R-\frac{1}{6}\right)^2 \right]\\
    &-\frac{m_{ph}^2}{16\pi^2}\left( \xi_R-\frac{1}{6} \right)G_{\mu\nu}+\frac{m_R^4}{64\pi^2}g_{\mu\nu}\Bigg{\}},
    \end{aligned}
\end{equation}
and the  divergent terms can be absorbed into the following  redefinition of the gravitational constants on the LHS of the SEE:

\begin{equation}
\begin{aligned}
    k_B^{-1}=k_R^{-1}+\frac{m_B^2}{8\pi^2}&\left\{ \left(\xi_R-\frac{1}{6} \right)\left[ \frac{1}{\epsilon}+\frac{1}{2}+\frac{1}{2}\ln\left( \frac{m_R^2}{\hat{\mu}^2} \right) \right]\right.\\
    &\left.-\left.\frac{dF_{dS}}{dR}\right|_{m_R^2,R_0} \right\},
    \end{aligned}
\end{equation}

\begin{equation}
\begin{aligned}
    &\Lambda_Bk_B^{-1}=\Lambda_Rk_R^{-1}\\
    &-\frac{m_B^2m_R^2}{32\pi^2}\left[ \frac{1}{\epsilon}+\frac{1}{2}\ln\left( \frac{m_R^2}{\hat{\mu}^2}\right)-\left.\frac{dF_{dS}}{dm_{ph}^2}\right|_{m_R^2,R_0} \right]-\frac{m_R^4}{64\pi^2},
    \end{aligned}
\end{equation}

\begin{equation}
\begin{aligned}
    &\alpha_{1B}=\alpha_{1R}\\
    &-\frac{(\xi_B-\frac{1}{6})}{16\pi^2}\left\{ \left(\xi_R-\frac{1}{6} \right)\left[ \frac{1}{\epsilon}+\frac{1}{2}+\frac{1}{2}\ln\left( \frac{m_R^2}{\hat{\mu}^2} \right) \right]\right.\\
    &\left.-\left.\frac{dF_{dS}}{dR}\right|_{m_R^2,R_0} \right\},
    \end{aligned}
\end{equation}

\begin{equation}
    \alpha_{2B}=\alpha_{2R}+\frac{1}{1440\pi^2}\left[ \frac{1}{\epsilon}+\frac{1}{2}+\frac{1}{2}\ln\left( \frac{m_R^2}{\hat{\mu}^2} \right) \right],
\end{equation}

\begin{equation}
    \alpha_{3B}=\alpha_{3R}-\frac{1}{1440\pi^2}\left[ \frac{1}{\epsilon}+\frac{1}{2}+\frac{1}{2}\ln\left( \frac{m_R^2}{\hat{\mu}^2} \right) \right].
\end{equation}

Hence, we can now write a finite expression for the SEE

\begin{equation}
\begin{aligned}
    &k_R^{-1}G_{\mu\nu}+\Lambda_Rk_R^{-1}g_{\mu\nu}+\alpha_{1R} ^{(1)}H_{\mu\nu}+\alpha_{2R} ^{(2)}H_{\mu\nu}+\alpha_{3R}H_{\mu\nu}\\
    &=\langle {T_{\mu\nu}}\rangle_{ren}+\langle {{T}_{\mu\nu}} \rangle_{ad4}^{con},
    \end{aligned}
\end{equation}
where $[\langle T_{\mu\nu} \rangle-\langle {{T}_{\mu\nu}} \rangle_{ad4}]=\langle {T_{\mu\nu}} \rangle_{ren}$.

Notice the above renormalization procedure only uses dS spacetime at the renormalization point. This is the main difference with respect to the traditional renormalization procedure for which a Minkowski spacetime is  used. 
The metric $g_{\mu\nu}$ involved in both sides of the  SEE (in the geometric tensors and in the stress energy tensor),  which is the solution of the SEE, is unspecified. The traditional equations are recovered in the MR scheme (i.e., when $R_0\to 0$).

\section{Renormalized semiclassical Einstein equations in de Sitter}\label{SecRenSEEdS}

Let us now specialize these results for de Sitter spacetimes. In dS, the geometric quantities appearing on the LHS of the SEE  are proportional to the metric $g_{\mu\nu}$, with a proportionality factor that depends on $R$ and the number of dimensions $d$

\begin{equation}
    \begin{aligned}
    R_{\mu\nu}&=\frac{R}{d}g_{\mu\nu},\\
    G_{\mu\nu}&=\left( \frac{1}{d}-\frac{1}{2} \right)Rg_{\mu\nu},\\
    ^{(1)}H_{\mu\nu}&=\frac{1}{2}\left(1-\frac{4}{d} \right)R^2g_{\mu\nu},\\
    ^{(2)}H_{\mu\nu}&=\frac{1}{2d}\left(1-\frac{4}{d} \right)R^2g_{\mu\nu},\\
    H_{\mu\nu}&=\frac{1}{d(d-1)}\left(1-\frac{4}{d} \right)R^2g_{\mu\nu}.
    \end{aligned}
    \label{TensoresDeSitter}
\end{equation}
Moreover, for any other tensor of  range two we have similar properties, for example,
\begin{equation}
    [G_{1;\mu\nu}]=\frac{1}{d}[\Box G_1]g_{\mu\nu}.
    \label{G1proporcionalagmunu}
\end{equation}
  de Sitter invariance also implies that every  scalar invariant is constant, and particularly $[G_1]$ is independent of the spacetime point.  Using this and (\ref{G1proporcionalagmunu}) in (\ref{ValorExpectTmunu2}),    it  is immediate to conclude that the tensor $\langle T_{\mu\nu}\rangle$  in Eq.~(\ref{ValorExpectTmunu})   is also proportional to $g_{\mu\nu}$  and  given by
\begin{equation}
    \begin{aligned}
    \langle T_{\mu\nu}\rangle&=Ng_{\mu\nu}\left[ -\frac{1}{2d}[\Box G_1]-\frac{m_B^2}{4}[G_1]\right.\\
    &\left. +\xi_B\frac{[G_1]}{2}\left(\frac{1}{d}-\frac{1}{2}\right)R-\frac{\lambda_B}{32}[G_1]^2\right].
    \end{aligned}
    \label{619Leo}
\end{equation} Using  
  the definition of $m_{ph}^2$ in Eq.~(\ref{Defmph}),

\begin{equation}
    \frac{\lambda_B}{4}[G_1]=m_{ph}^2-\delta\overline{\xi}R-m_B^2,
\end{equation} we obtain

\begin{equation}
    \begin{aligned}
    &\langle T_{\mu\nu}\rangle =Ng_{\mu\nu}\left\{ \frac{\xi_B R}{2(4+\epsilon)}\left[\frac{4}{\lambda_B}(m_{ph}^2-\delta\overline{\xi}R-m_B^2) \right]\right.\\
    &\left.-\frac{1}{2}(m_{ph}^2+\xi_R R)\right.\\
     & +\left.\frac{\lambda_B}{32}\left[ \frac{4}{\lambda_B}(m_{ph}^2-\delta\overline{\xi}R-m_B^2) \right]^2\right\},
    \end{aligned}
\end{equation}where  $\epsilon=d-4$.
Notice we cannot yet take $\epsilon\to0$ in the denominator, due to the fact that it is multiplied by bare parameters. In order  to perform such a limit, first we need to remind the expressions (\ref{lambdaRR0cte}), (\ref{lambdaRestrella}) and (\ref{RelacionUtilxiRlambdaR}). After some algebra and after neglecting the $\mathcal{O}(\epsilon)$ terms, we obtain

\begin{equation}
    \begin{aligned}
    &\langle T_{\mu\nu}\rangle=Ng_{\mu\nu}\left\{\frac{1}{128\pi^2}\left[ m_{ph}^2+\left(\xi-\frac{1}{6}\right)R \right]^2\right.\\
    &+\frac{1}{2}\left[\delta\overline{m}^2+\left(1+\frac{\epsilon}{4+\epsilon}\right)\delta\overline{\xi}R\right]\left(\frac{m_R^2}{\lambda_R^*}+\frac{(\xi_R-\frac{1}{6}R_0)}{32\pi^2}\right)\\
    &+\left(\frac{4}{4+\epsilon}\right)\frac{\epsilon\delta\overline{\xi}}{8}\left(\frac{3(\xi_R-\frac{1}{6})}{\lambda_R}+J\right)R^2\\
    &\left. \frac{1}{2}\left(\frac{m_R^2}{\lambda_R^*}+\frac{(\xi_R-\frac{1}{6})R_0}{32\pi^2}\right)(m_R^2-m_{ph}^2) \right\}
    \end{aligned}
\end{equation}

To compute the renormalized expectation value, $\langle T_{\mu\nu} \rangle_{ren}=\langle T_{\mu\nu} \rangle-\langle T_{\mu\nu} \rangle_{ad4}$, we must evaluate $\langle T_{\mu\nu} \rangle_{ad4}$ from Eq. \eqref{Tmunuad4} for the de Sitter spacetime. It is then when we use the expressions in Eq. \eqref{TensoresDeSitter}. We now use that $\langle {T_{\mu\nu}}\rangle_{ad4}=\langle {T_{\mu\nu}}\rangle_{ad4}^{div}+\langle {T_{\mu\nu}}\rangle_{ad4}^{con}$, where

\begin{equation}
    \begin{aligned}
    &\langle  {T_{\mu\nu}} \rangle_{ad4}^{con}=N g_{\mu\nu}\left\{\frac{m_R^2}{2}\left[ \frac{m_R^2}{\lambda_R^*}+\frac{(\xi_R-\frac{1}{6})R_0}{32\pi^2}+\frac{m_R^2}{32\pi^2} \right]\right.\\
    &+\frac{m_{ph}^2}{64\pi^2}\left(\xi_R-\frac{1}{6} \right)R-\frac{m_{ph}^2}{32\pi^2}R\left.\frac{dF_{dS}}{dR}\right|_{m_R^2,R_0}\\
     &+\frac{m_{ph}^4}{64\pi^2}\left[ \frac{32\pi^2}{\lambda_R^*}+\frac{1}{2}+\frac{(\xi_R-\frac{1}{6})R_0}{m_R^2}-2\left.\frac{dF_{dS}}{dm_{ph}^2}\right|_{m_R^2,R_0} \right]\\
      &-\frac{m_{ph}^2}{64\pi^2}\left[ m_{ph}^2+\left( \xi_R-\frac{1}{6} \right)R \right]\ln\left( \frac{m_{ph}^2}{m_R^2} \right)\\
      &\left. -m_{ph}^2\left[ \frac{m_R^2}{\lambda_R^*}+\frac{(\xi_R-\frac{1}{6})R_0}{32\pi^2} \right]\right\}
    \end{aligned}
    \label{Tmunuad4conDS}
\end{equation}

\begin{equation}
    \begin{aligned}
    &\langle {T_{\mu\nu}}\rangle_{ad4}^{div}=N g_{\mu\nu}\left\{ \frac{1}{64\pi^2}\frac{R^2}{2160}-\frac{m_R^4}{64\pi^2}\right.\\
    &+\left(\frac{4}{4+\epsilon}\right)\frac{\epsilon\delta\overline{\xi}}{8}\left(\frac{3(\xi_R-\frac{1}{6})}{\lambda_R}+J\right)R^2  \\
    &\left.+\frac{1}{2}\left[\delta\overline{m}^2+\left(1+\frac{\epsilon}{4+\epsilon}\right)\delta\overline{\xi}R\right]\left(\frac{m_R^2}{\lambda_R^*}+R_0\frac{(\xi_R-\frac{1}{6})}{32\pi^2}\right)\right\}
    \end{aligned}
\end{equation}

Then, after subtracting    the tensor $\langle {T_{\mu\nu}} \rangle_{ad4}$ given in Eq.~(\ref{Tmunuad4}),  and neglecting the terms that are $\mathcal{O}(\epsilon)$, the result can be written as

\begin{equation}
    \begin{aligned}
    &\langle T_{\mu\nu}\rangle_{ren}=-N\frac{g_{\mu\nu}}{64\pi^2}\left\{m_{ph}^2\left[ \left( \frac{32\pi^2}{\lambda_R^*}+\frac{(\xi_R-\frac{1}{6})R_0}{m_R^2}\right.\right.\right.\\
    &\left.\left.\left.-2\left.\frac{dF_{dS}}{dm_{ph}^2}\right|_{m_R^2,R_0} \right)(m_{ph}^2-m_R^2) \right]\right.\\
     &-2m_{ph}^2\left[ R\left.\frac{dF_{dS}}{dR}\right|_{m_R^2,R_0}+m_{R}^2\left.\frac{dF_{dS}}{dm_{ph}^2}\right|_{m_R^2,R_0} \right]\\
     &-\frac{1}{2}\left(\xi_R-\frac{1}{6} \right)^2R^2+\frac{R^2}{2160}\\
      &-\left.m_{ph}^2\left[ m_{ph}^2+\left(\xi_R-\frac{1}{6} \right)R \right]\ln\left( \frac{m_{ph}^2}{m_R^2}\right)\right\}.
    \end{aligned}
    \label{TmunurenDS}
\end{equation}

Therefore, the RHS of the SEE in de Sitter is given by

\begin{equation}
    \begin{aligned}
    &\langle T_{\mu\nu}\rangle_{ren}+\langle {T_{\mu\nu}}\rangle_{ad4}^{con}=\\
    &-Ng_{\mu\nu}\frac{1}{64\pi^2}\left\{ 32\pi^2\left( \frac{m_R^2}{\lambda_R^*}+\frac{(\xi_R-\frac{1}{6})R_0}{32\pi^2} \right)(m_{ph}^2-m_R^2) \right.\\
     &-\left.m_R^4+\frac{R^2}{2160}-\frac{1}{2}\left[ m_{ph}^2+\left(\xi_R-\frac{1}{6}\right)R \right]^2\right\}.
    \end{aligned}
    \label{RHS_EES_DS}
\end{equation}

\noindent The  LHS is simply

\begin{equation}
   k_R^{-1} G_{\mu\nu}+\Lambda_R k_R^{-1}g_{\mu\nu}=k_R^{-1}\left( -\frac{R}{4}+\Lambda_R \right)g_{\mu\nu}.
    \label{LHS_EES_DS}
\end{equation}
\noindent
The quadratic tensors that were introduced, $^{(1)}H_{\mu\nu}$, $^{(2)}H_{\mu\nu}$ and $H_{\mu\nu}$, vanish for $d=4$. However, their presence was important for the renormalization procedure (and  there is a finite remnant on the RHS of the SEE due to the well-known trace anomaly). 
Then, as seen in Eq. (\ref{RHS_EES_DS}) and in Eq. (\ref{LHS_EES_DS}), we can factorize the metric $g_{\mu\nu}$ from both sides, obtaining a scalar and algebraic equation with a sole degree of freedom of the metric, $R$,

\begin{equation}
    \begin{aligned}
    &8\pi M_{pl}^2\left( -\frac{R}{4}+\Lambda_R \right) =-\frac{R^2}{2160}+\frac{1}{2}\left[ m_{ph}^2+\left( \xi_R-\frac{1}{6} \right)R \right]^2\\
     &-32\pi^2\left( \frac{m_R^2}{\lambda_R^*}+\frac{(\xi_R-\frac{1}{6})R_0}{32\pi^2} \right)(m_{ph}^2-m_R^2)+m_R^4
    \end{aligned}
    \label{EES_DS_Renormalizadas}
\end{equation}

\noindent where we have divided both sides by $N$ and defined a rescaled Planck mass  $M_{pl}$, that respects $N k_R =8\pi/M_{pl}^2=8\pi G_N$.

\section{De Sitter self-consistent solutions}\label{SecSelfC}
   
 We are now ready to study the backreaction effect on the spacetime curvature due to the computed  quantum corrections. In Sec. II we found that a dynamical physical mass $m_{ph}$ is generated, which obeys Eq.~(\ref{mphR0cte}) with $\hat{\phi}=0$.
Notice that $m_{ph}$ depends not only  on the curvature scalar $R$, but also  on the parameters $m_R$, $\xi_R$ $\lambda_R$, and the  curvature $R_0$ defined at the renormalization point. In Sec. III, the renormalized SEE were derived, which reduce to the algebraic equation  for $R$ given in Eq. (\ref{EES_DS_Renormalizadas}).

Our goal in this section is to solve both Eq. (\ref{mphR0cte}), in the symmetric phase ($\hat{\phi}=0$), and Eq. (\ref{EES_DS_Renormalizadas})  self-consistently, for different renormalization schemes and values of the free parameters. We do this numerically. First, in Sec.\ref{subphysM}, we analyze the effects on the physical mass characterizing its departures from the renormalized value $m_R$. Then, in Sec.\ref{subphysM}, we concentrate on the effects on the spacetime curvature $R$, which can be characterized as departures from the classical solution  $(R-R_{cl})/{R_{cl}}$ and/or from its value at the renormalization point $(R-R_0)/{R_0}$. We focus on the infrared limit,  namely $m_{ph}^2+\xi_RR\ll R$, so that, after using Eqs. \eqref{FdS} and \eqref{gLimIR},  Eq. (\ref{mphR0cte})  can be approximated  by 
\begin{equation}
    \begin{aligned}
    &m_{ph}^2+\xi_RR=m_R^2+\xi_RR\\
    &+\frac{\lambda_R^*}{32\pi^2}\left\{\frac{R^2}{24(m_{ph}^2+\xi_RR)}-\xi_RR-\frac{5R}{36}\right.\\
     &-\frac{R}{6}[\kappa+\log(R/12m_R^2)]\\
     &-(m_R^2+\xi_RR)\left[2\left.\frac{dF_{dS}}{dm^2}\right|_{m_R^2,R_0}-\frac{(\xi_R-\frac{1}{6})R_0}{m_R^2}\right]\\
      & +2\left( F_{dS}(m_R^2,R_0)+\left. \frac{dF_{dS}}{dR} \right|_{m_R^2,R_0}(R-R_0) \right)\\
       & +\left.\left[ \kappa+\log(R/12m_R^2)-\frac{46}{54}+2\left.\frac{dF_{dS}}{dm^2}\right|_{m_R^2,R_0}\right.\right.\\
       &\left.\left.-\frac{(\xi_R-\frac{1}{6})R_0}{m_R^2} \right](m_{ph}^2+\xi_RR)\right\},
    \end{aligned}
    \label{mph_MasaPequena}
\end{equation}
where we remind that $\kappa=11/6-2\gamma_E$. This is a quadratic equation for $(m_{ph}^2+\xi_RR)/R$ 

\begin{equation}
    A_{dS}\left\{\frac{(m_{ph}^2+\xi_RR)^2}{R^2}\right\}+B_{dS}\left\{\frac{(m_{ph}^2+\xi_RR)}{R}\right\}+C_{dS}=0,
\end{equation}
where the functions $A_{dS}$, $B_{dS}$ and $C_{dS}$ are defined as:

\begin{equation}
    \begin{aligned}
    &A_{dS} =1-\frac{\lambda_R^*}{32\pi^2}\left[ \kappa+\log(R/12m_R^2)-\frac{49}{54}\right.\\
    &\left.+2\frac{dF_{dS}}{dm^2}|_{m_R^2,R_0}-\frac{(\xi_R-\frac{1}{6})R_0}{m_R^2} \right],\\
   & B_{dS}=-\left(\frac{m_R^2}{R}+\xi_R \right)+\frac{\lambda_R^*}{32\pi^2}\left\{ \frac{\kappa}{6}+\frac{1}{6}\log(R/12m_R^2)\right.\\
     & +\frac{5}{36}+\frac{(m_R^2+\xi_RR)}{R}\left[ 2\left. \frac{dF_{dS}}{dm^2} \right|_{m_R^2,R_0}-\frac{(\xi_R-\frac{1}{6})R_0}{m_R^2} \right]\\
      & \left.+\xi_R-\frac{2}{R}\left[ F_{dS}(m_R^2,R_0)+\left. \frac{dF_{dS}}{dR} \right|_{m_R^2,R_0}(R-R_0) \right]\right\},\\
   &  C_{dS}=-\frac{\lambda_R^*}{768\pi^2}.\nonumber
    \end{aligned}
\end{equation} 
So, the physical mass $m_{ph}^2$ in the infrared limit can be expressed as
\begin{equation}
    m_{ph}^2=\frac{-RB_{dS}\pm\sqrt{(RB_{dS})^2-4R^2A_{dS}C_{dS}}}{2A_{dS}}-\xi_RR\,,
    \label{mph_Cuadratica}
\end{equation} which is consistent with the results previously obtained in \cite{RevisitedI}. 
%For $\lambda_R\ll 1$, by expanding this result at leading order in $\sqrt{\lambda}$  it is straightforward to s we 

\subsection{$m_{ph}^2$ analysis}\label{subphysM}

First of all, we  need to make sure that the parameters of the problem  ensure that the quantity
$m_{ph}^2$ is real and positive, otherwise, the de Sitter-invariant propagator solution would not be valid. 
Besides it is necessary to  recall we are restricting  the analysis  to the infrared regime, meaning $m_{ph}^2+\xi_RR\ll R$. We define the variable 

\begin{equation}\label{VariabDelta}
    \Delta_{m^2}=\frac{m_{ph}^2-m_R^2}{m_R^2}.
\end{equation} Recall the renormalized mass is, by definition, the physical mass evaluated at the renormalization point. So, this  variable   characterizes  how much the physical mass departs from the renormalized one when  the  values of the curvature are beyond that point (which is generically the case if $R_0$ is not the full solution of the problem including quantum effects -which is unknown beforehand-).  Assuming $\xi_R\ll 1$, the condition $m_{ph}^2+\xi_RR\ll R$ can be replaced by  $m_{ph}^2\ll R$. In order to check whether this condition is  satisfied, in what follows we use the following inequality: $m_{ph}^2\leq R/10$, or equivalently \footnote{The factor $1/10$ roughly corresponds to an order $1\%$ error in the approximation Eq. (\ref{gLimIR}) of the function $F_{dS}$ defined in Eq. (\ref{FdS}). }
\begin{equation}
    \Delta_{m^2}\leq\Delta_{c^2}\equiv\frac{1}{10}\frac{R}{m_R^2}-1.
    \label{RestriccionMasa} 
\end{equation}

Throughout this section we study three cases:   the $R_0=0$ case that corresponds to a flat curvature or a Minskowski spacetime, the case  $R_0>0$ with $R_0= 10^{-6}M_{pl}^2$ or  $R_0= 10^{-2}M_{pl}^2$ as examples of  possible  values for a curved spacetime in the semiclassical regime. One can see that the results do not depend strongly on the particular value of $R_0$ as long as $R_0\ll M_{pl}^2$, and   the case $R_0=R_{cl}$, where $R_{cl}$ is the solution to  the background curvature  with the quantum effects  neglected.  We present the plots for  two values of the renormalized mass, corresponding to  $m_R^2=10^{-7}M_{pl}^2$  and  $m_R^2=10^{-3}M_{pl}^2$.  
For the case $R_0>0$, we take 
   $R_0= 10^{-6}M_{pl}^2$ for the  smaller mass and we focus on  $R_0= 10^{-2}M_{pl}^2$ for the larger mass. We make this choice for 
 $m_R^2=10^{-7}M_{pl}^2$  because whereas for $R_0= 10^{-6}M_{pl}^2$  the physical mass is well defined (it is real and positive) for all values of the parameters and range of $R$  we study, for $R_0= 10^{-2}M_{pl}^2$ this is not always true if $R<R_0$. For $m_R^2=10^{-3}M_{pl}^2$ the physical mass is well defined for both values  $R_0= 10^{-6}M_{pl}^2$ and $R_0= 10^{-2}M_{pl}^2$, but for $R_0= 10^{-2}M_{pl}^2$ the curves can  be easily distinguished from the ones with $R_0=0$.  
 
\begin{widetext}

\begin{figure}[H]

\begin{subfigure}[b]{.49\linewidth}
\includegraphics[width=\linewidth]{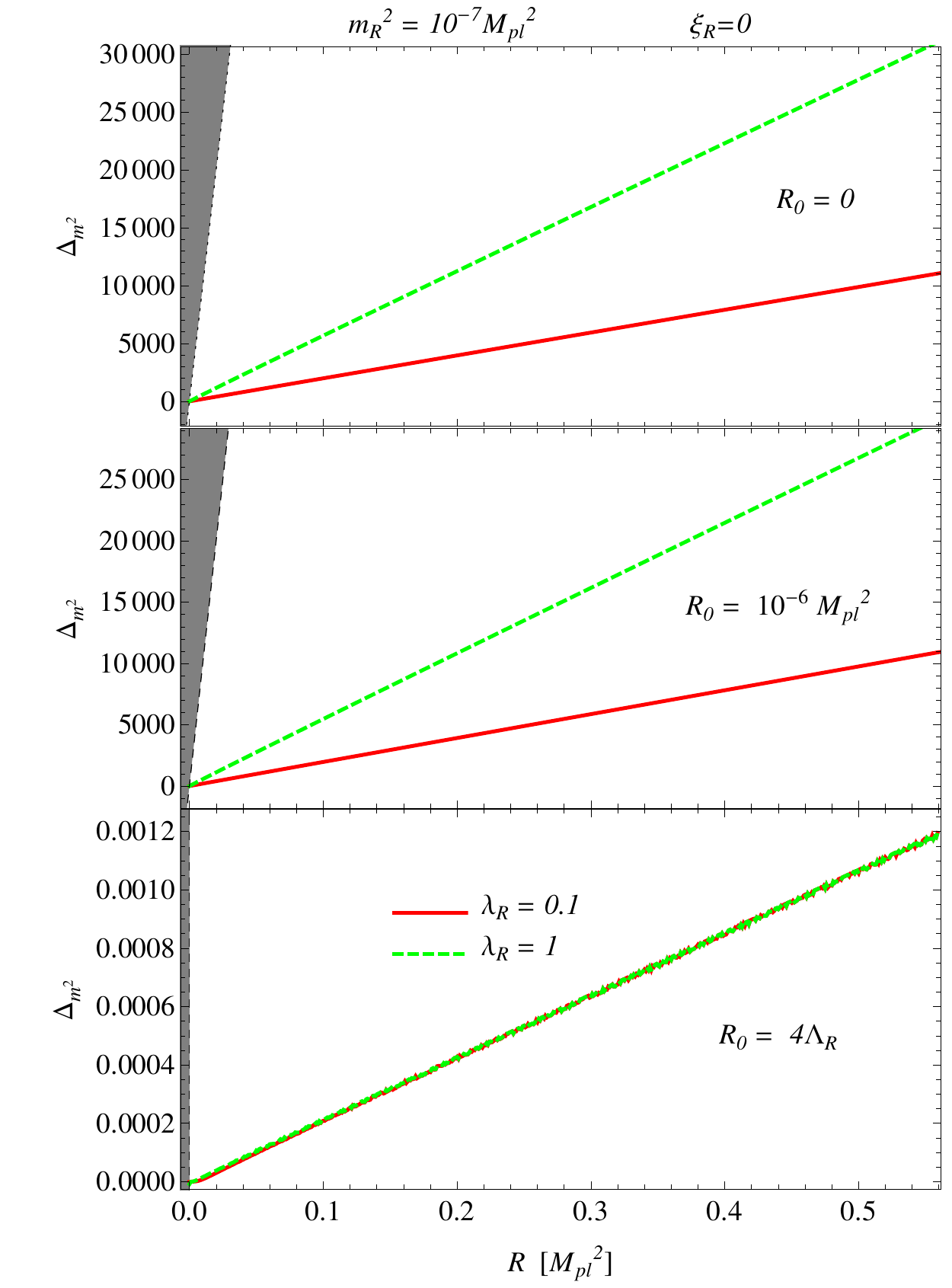}
\end{subfigure}
\begin{subfigure}[b]{.49\linewidth}
\includegraphics[width=\linewidth]{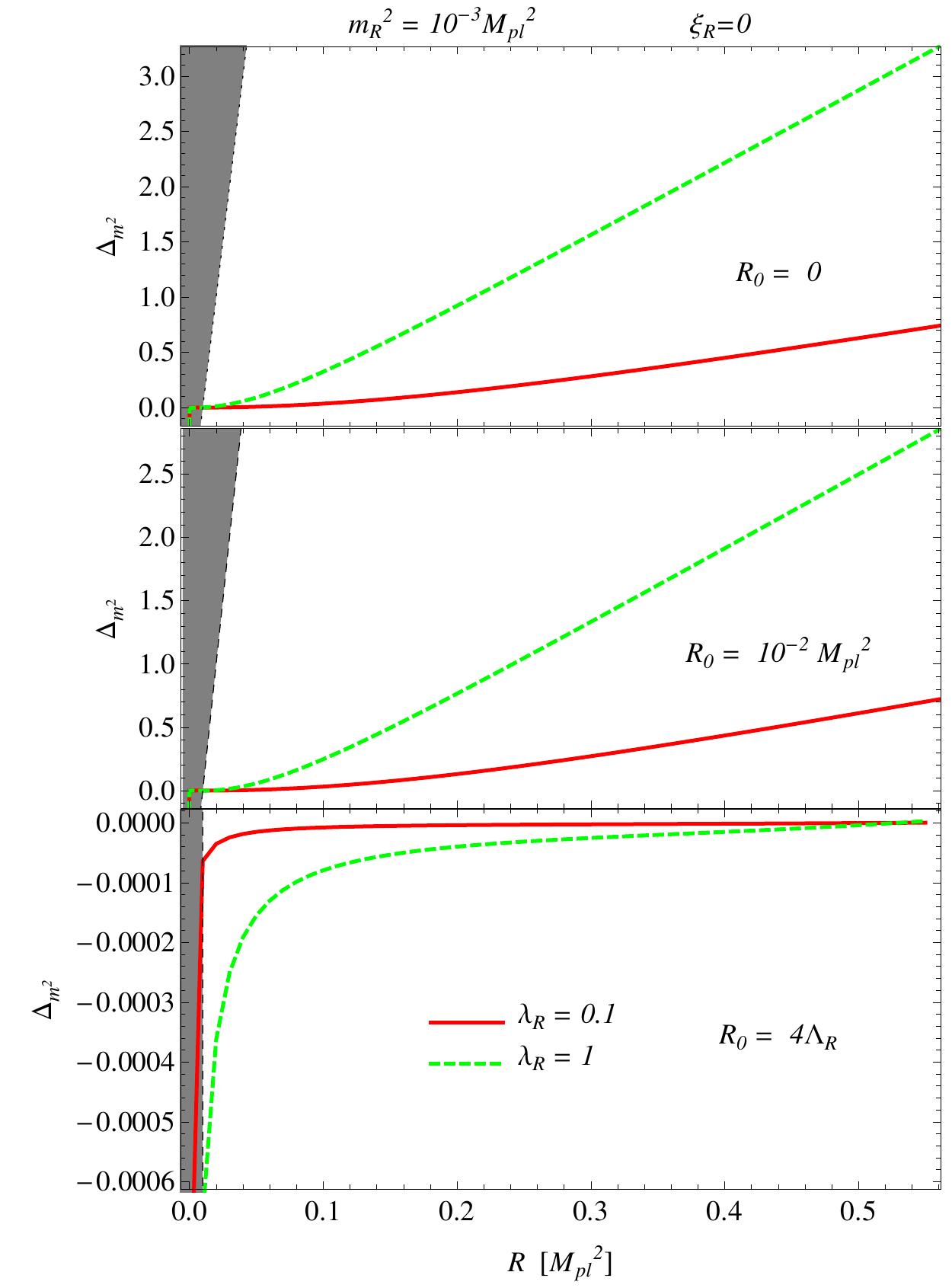}
\end{subfigure}

\caption{The variable $\Delta_{m^2}$ defined in Eq. (\ref{VariabDelta}) as a function of $R$, for $\xi_R=0$ and two different values of $\lambda_R$: $\lambda_R=0.1$ (green dashed lines) and $\lambda_R=1$ (red solid lines). The first column corresponds to $m_R^2=10^{-7}M_{pl}^2$ and the second one to $m_R^2=10^{-3}M_{pl}^2$. The first row is for $R_0=0$. In the second row the case  $R_0= 10^{-6}M_{pl}^2$ is shown for the smaller mass, whereas for the larger mass the curve is for $R_0=10^{-2}M_{pl}^2$. The last row is for $R_0=4\Lambda_R$. The grey area corresponds to values for which the  restriction  $\Delta_{m^2}<\Delta_{c^2}$ is violated [see Eq. (\ref{RestriccionMasa})]. On the left   panel, the two curves in the  bottom plot are almost superimposed and cannot be distinguished by eye.}
\label{mph1to4}
\end{figure}

\end{widetext} 

From   Fig. \ref{mph1to4} one can see that for the same values of the curvature $R$ (shown on the common horizontal axes) the order of magnitude of the vertical axes change significantly depending on $R_0$ (the value of $R$ at the renormalization point). For intermediate values of the parameters the obtained results are similar and lay between the corresponding curves.  The departures characterized by $\Delta_{m^2}$ are significantly larger for lighter fields, as can be seen by comparing the right panel (which corresponds to $m_R^2=10^{-3}M_{pl}^{2}$) to the left one (where  $m_R^2=10^{-7}M_{pl}^2$).  This means that  when the fields are  light, the physical mass is less robust against  corrections beyond the
renormalization point. This result is an expected manifestation of the infrared sensitivity of light fields to the spacetime curvature.

\begin{widetext}

\begin{figure}[H]
\begin{subfigure}[b]{.49\linewidth}

\includegraphics[width=\linewidth]{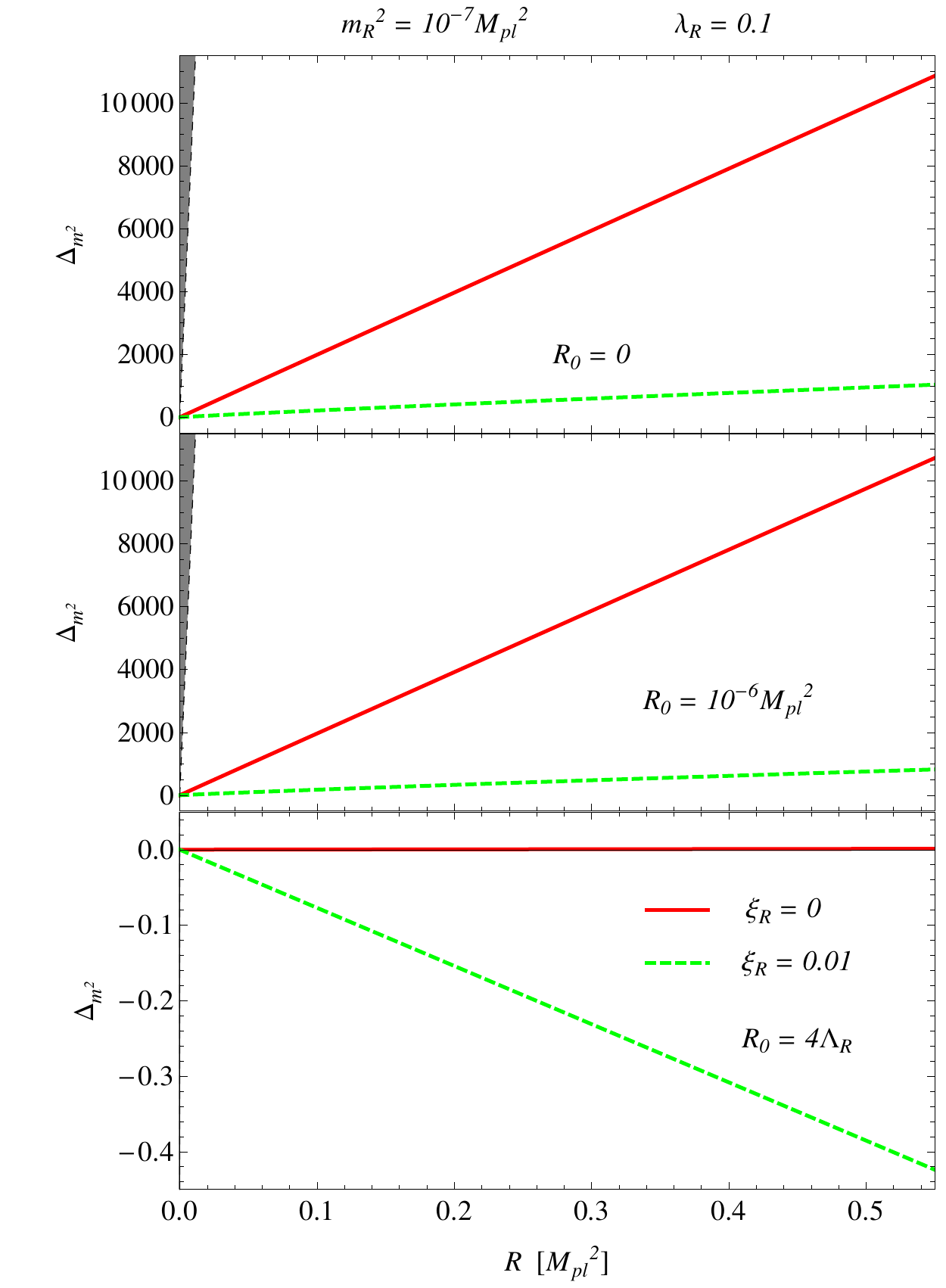}\end{subfigure}\begin{subfigure}[b]{.49\linewidth}
\includegraphics[width=\linewidth]{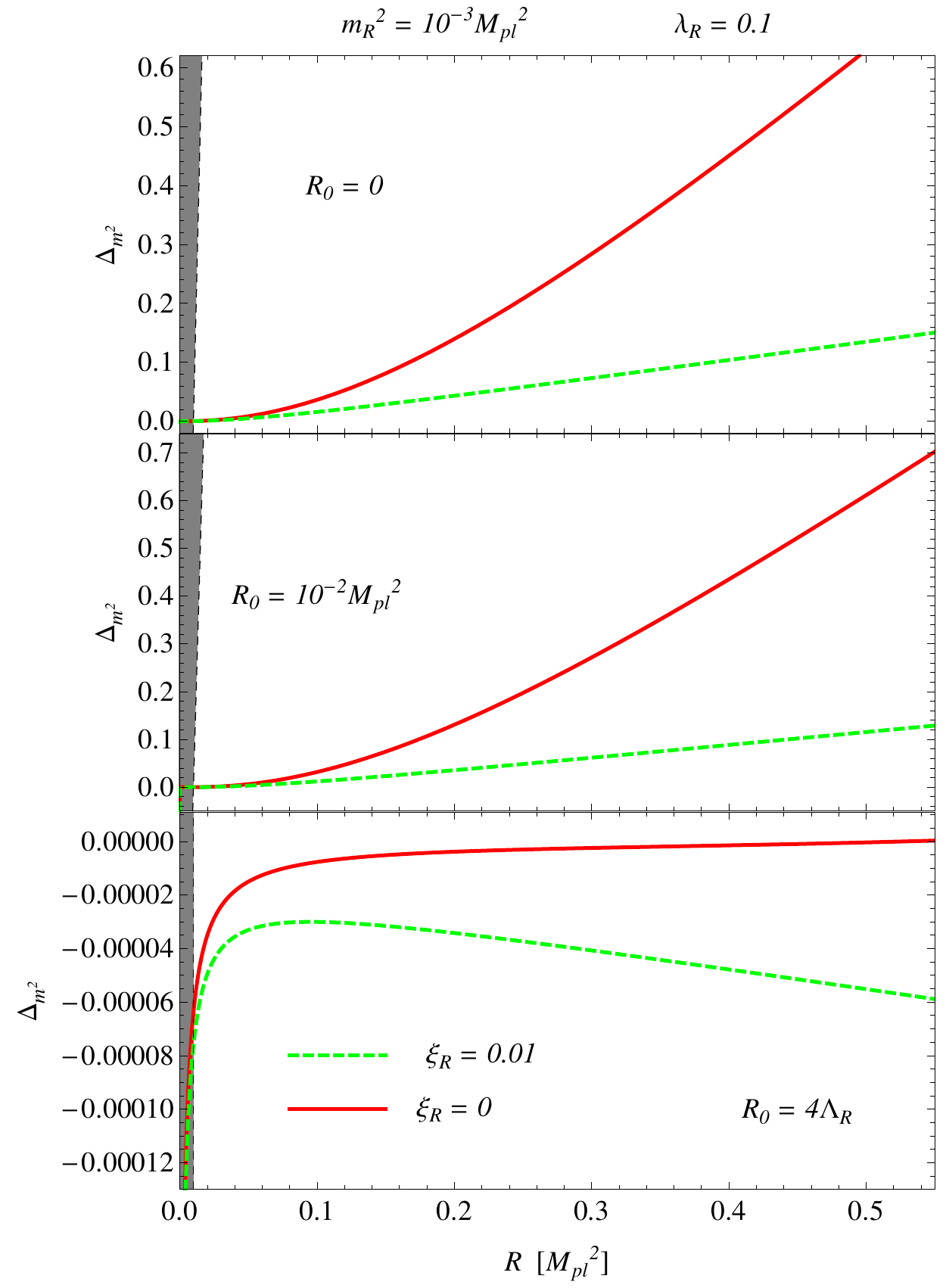}\end{subfigure}
 \caption{$\Delta_{m^2}$ vs $R$, where we fixed $\lambda_R=0.1$, for $m_R^2=10^{-7}M_{pl}^2$ (on the left) and  $m_R^2=10^{-3}M_{pl}^2$ (to the right). We show the cases $R_0=0$ (top), $R_0=0.01M_{pl}^2$ (middle) and $R_0=4\Lambda_R$ (bottom). $\xi_R$ was varied between the values $\xi_R=0$ (solid red lines) and $\xi_R=0.01$ (green dotted lines). The region for which   $\Delta_{m^2}<\Delta_{c^2}$  is shown as a grey-painted area.}
\label{mph7mph8mph9}
\end{figure}
\end{widetext}
 From  Fig. \ref{mph1to4} one can also see that $|\Delta_{m^2}|$ is larger the larger the coupling constant  $\lambda_R$. For the  plots on the top and in the middle, since $m_R^2$ does not depend on $R$, this means that  when the interaction between the scalar fields intensifies, the physical mass becomes more sensitive to  changes of the curvature. This can also be seen analytically  from Eq. (\ref{mphR0cte}).  Indeed, under the same condition we are assuming to make the plots, $m^2_{ph}+\xi_R R\ll R$, using Eqs. (\ref{FdS}) and Eq .(\ref{g}), from Eq. (\ref{mphR0cte}) one can immediately conclude that the physical mass scales almost linearly with the curvature. Using the same formulas one can also see that the dependence on $\lambda_R$ is suppressed when the physical mass is close to the renormalized one and $R$ is close to $R_0$.  Notice the part of  $F_{dS}$ that is linear in $R$ cancels out in the last square bracket of Eq. (\ref{mphR0cte}).  
For the plots on the bottom, the interpretation  is not so  straightforward. 
This is because the renormalization point is fixed by the classical background curvature, $R_0=R_{cl}$, which is determined by the cosmological constant, $R_{cl}= 4 \Lambda_R$.  For each value  of $R$ on the horizontal axes, the value of $\Lambda_R$ is such that   the quantum corrections to $R_{cl}$  yield such value of $R$ as the self-consistent solution of the gap equation (\ref{mphR0cte}) and the SEE (\ref{EES_DS_Renormalizadas}). Therefore, the  value of $\Lambda_R$ is different for each point in the curves. So, for such plots, the renormalized mass defined at $R_0=4\Lambda_R$  can be thought as a function of $R$. Notice that taking into account  that the physical value of  $R$ is obtained by  solving  the renormalized SEE self-consistently, the corresponding value of $\Lambda_R$ is also different for each point in the curves drawn  in the plots at the top and in the middle of Fig.~\ref{mph1to4}. In other words, in those plots the value of   $\Lambda_R$ is assumed to be independent of $R_0$, so by properly choosing $\Lambda_R$ one can make all values of $R$ to correspond to a solution of the SEE. For this reason it is not  necessary to use explicitly the SEE to make those plots.

The use of the  scheme with $R_0=4\Lambda_R$, where  the parametrization of the theory is done in the classical background metric, has remarkable  properties.  In particular, for the scheme with $R_0=4\Lambda_R$,  if the  IR approximation is valid  for $m_R^2$ ($m_R^2+\xi_R R\ll R$) it remains to be valid also for $m_{ph}$ ($m_{ph}^2+\xi_R R\ll R$), even for large values of the coupling constant. This is a nice property of the scheme, since in practice, when studying the backreaction problem  the approximation is in general  very  useful.
 As we argue below, these   properties  make this case  ultimately the most convenient to study the self-consistent solutions and to assess the quantum backreaction effects produced by the quantum fields. 

Let us now study what happens to $\Delta_{m^2}$ when $\xi_R$ varies. 
  Figure \ref{mph7mph8mph9} shows  plots of $\Delta_{m^2}$ as a function of $R$, where we fixed   $\lambda_R=0.1$ for  $m_R^2=10^{-7}M_{pl}^2$ (on the left) and  $m_R^2=10^{-3}M_{pl}^2$ (to the right). The sensibility of $\Delta_{m^2}$ against changes of $R$ turns out to be minimal when setting $R_0=R_{cl}$. 
A similar conclusion to the previous cases can be drawn when we vary $\lambda_R$. 

\subsection{Backreaction solutions}\label{subBack}

Once the limits upon the physical mass in  Eq.~(\ref{RestriccionMasa}) are established, we can proceed to find the $\Lambda_R$ values, by solving self-consistently the system formed by the gap equation for $m_{ph}^2$ Eq. (\ref{mphR0cte}), with $\hat{\phi}=0$, and the SEE  (\ref{EES_DS_Renormalizadas}). To measure the departures of the scalar curvature from the classical one, we use the variable

\begin{equation}
    \Delta_R=\frac{R-R_{cl}}{R_{cl}}=\frac{R-4\Lambda_R}{4\Lambda_R}
    \label{DeltaR}.
\end{equation}
In what follows, we present plots of  $\Delta_R$    as a function of $\Lambda_R$, for different values of the parameters.  

    In order to interpret the plots it is useful to have in mind that the backreaction of the quantum fields depends on the renormalized parameters and $R_0$  as shown explicitly in Eq.~(\ref{EES_DS_Renormalizadas}). There is always a scale to be fixed in addition to the renormalized parameters associated to the bare parameters of the theory (in the MS scheme the scale is $\hat{\mu}$, whereas in the  other schemes we are considering it is $R_0$). For a given value of $m_R$, when the physical mass is close to   $m_R$, as can be seen from the right-hand side (rhs) of Eq.~(\ref{EES_DS_Renormalizadas}), the $\lambda_R$ and $R_0$ dependence are suppressed. From there, one can also see that for sufficiently large values of $R$ all solutions should approach to each other.

    In our study, we are restricting to  IR fields, meaning $m_{ph}^2+\xi_R R\ll R$, but notice that we have not   used such assumption to arrive at Eq.~(\ref{EES_DS_Renormalizadas}). Here we make use of the IR  approximation with $\xi_R\ll 1$  to compute the physical mass. Therefore, it is necessary to take into account the  upper limit  on the physical mass obtained from the analysis of the previous section [see Eq.~(\ref{RestriccionMasa}].  In this case, this   shows up as a  restriction  for   the possible values that $R$ can take, which corresponds to a  lower limit on $R$ that we call $R_{min}$. In order to indicate the part of the curves for which the restriction is not valid, we use grey dotted lines, and solid or dashed lines otherwise.

 The top plot of Fig. \ref{Back3Back5A} shows    $\Delta_R$ vs $\Lambda_R$, for $\lambda_R=0.1$, $\xi_R=0$, for the three $R_0$ cases  we are considering and two values of the mass: $m_R^2=10^{-7}M_{pl}^2$ (on the left) and $m_R^2=10^{-3}M_{pl}^2$ (to the right). 
 For  the heavier case, we obtain relatively larger variations with respect to the renormalization point. A relatively strong $R_0$ dependence is expected as  the absolute difference between the physical mass and the renormalized mass is larger.  On the other hand, for $R_0=R_{cl}$, since the physical mass is close to the renormalized one for all plotted values of $R$ (we are restricting to sub-Planckian values), the $R_0$  dependence is suppressed.   For the higher mass case,  in the plotted (sub-Planckian)  range of $R$,  the quantity $R-R_{cl}$ is only negative for the case $R_0=4\Lambda_R$. 
On the contrary,   for the lighter   case  $R-R_{cl}$ is in all cases negative, meaning  that the curvature scalar that  includes the quantum effects  is smaller than the classical one. For the smaller mass, the violet  curves cannot be distinguished  by eye because they are almost superimposed with the green ones. It can be seen the green and  violet curves depart from the others more as $R$ increases, which  is also compatible with the fact that, for those values of $R$, $m_{ph}$  becomes significantly different from $m_R$ for such schemes.  For $R_0=0$ and $R_0=10^{-2} M_{pl}^2$ in the heavier case, it can be shown that  if $R$ is allowed to take  super-Planckian  values,  as $\Lambda_R$ increases the curves go down  and approach to each other, obtaining  also $R<R_{cl}$ in such regime. This  indicates   the screening of the classical solution is  recovered in the IR regime. This result, $R<R_{cl}$, is obtained for smaller values of $R$ when the coupling is larger, as can be seen from the plot in the middle, where the case   $\lambda_R=1$ is also shown. The $\lambda_R$ dependence is stronger in the cases $R_0=0$ and fixed $R_0>0$ than   for $R_0=R_{cl}$. In view of Fig. 1, this result is consistent with what we have concluded above from  Eq.~(69). 
For the lighter case, no  significant variations are obtained, whereas once again $R-R_{cl}$ turns out to be negative.
  In the plot on the bottom of Fig. 3, the curves for $\xi_R=0.01$ are shown in addition to the ones for $\xi_R=0$. One can see that the absolute value of $\Delta_R$ is smaller for $\xi_R=0.01$, which can be expected from the arguments above and  Fig. \ref{mph7mph8mph9}. Notice for the lighter case, all curves with $\xi_R=0.01$ are indistinguishable by eye.

\begin{widetext}

\begin{figure}[H]

\begin{subfigure}[b]{.49\linewidth}
\includegraphics[width=\linewidth]{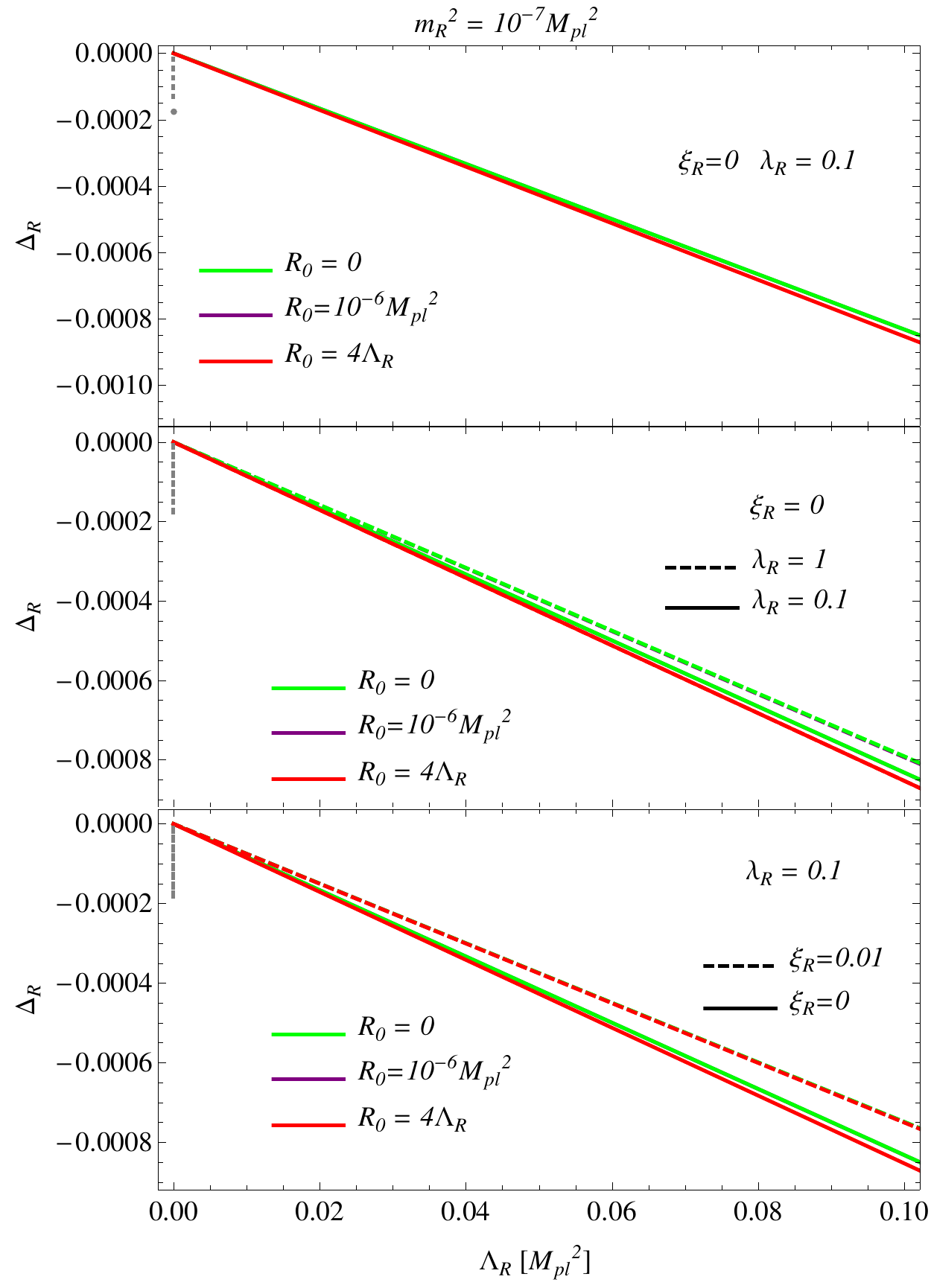}
\end{subfigure}
\begin{subfigure}[b]{.49\linewidth}
\includegraphics[width=\linewidth]{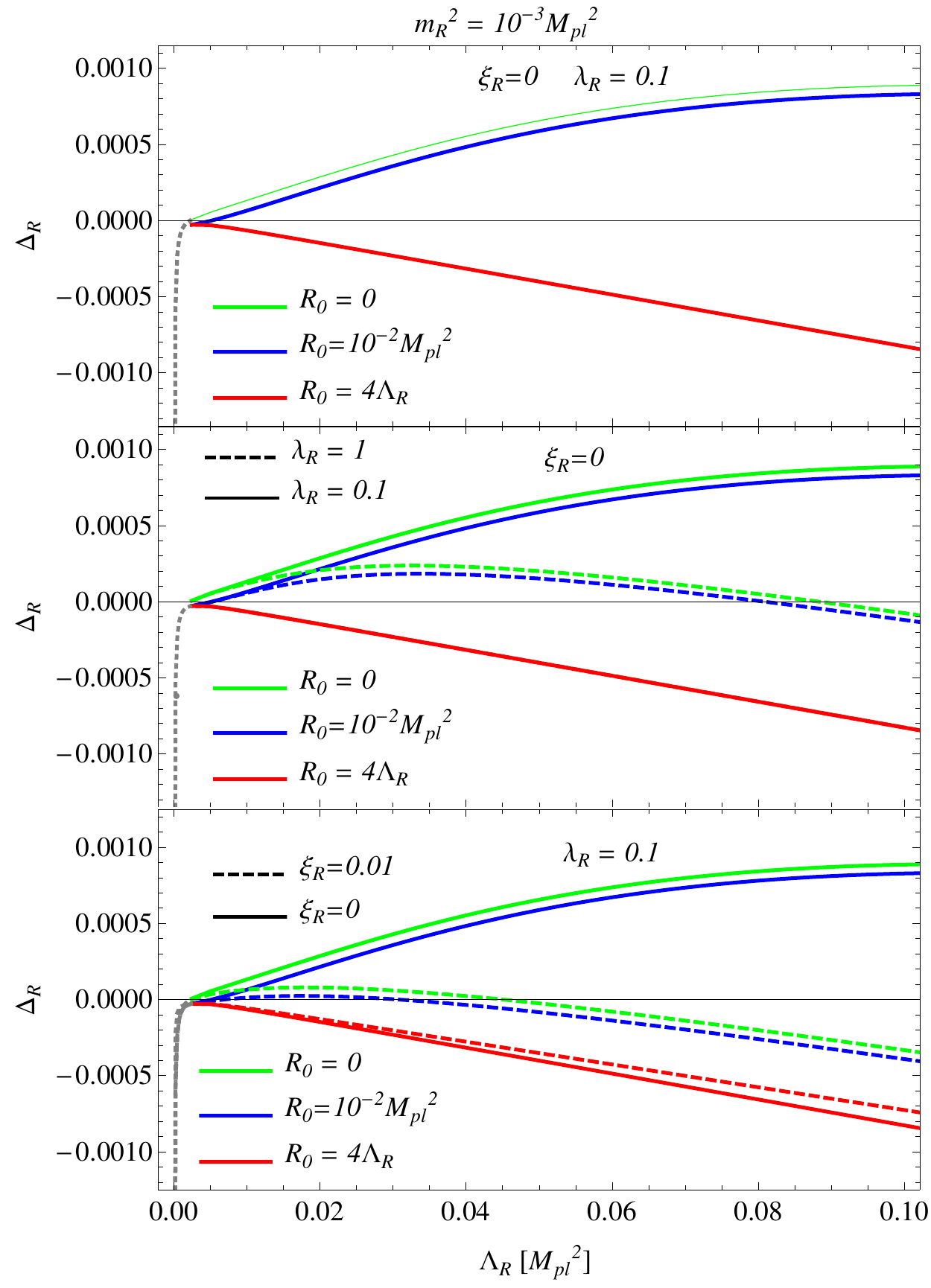}
\end{subfigure}
\caption{  $\Delta_R=\frac{R-R_{cl}}{R_{cl}}$ vs $\Lambda_R$. The plots on the right correspond to the $m_R^2=10^{-3}M_{pl}^2$ case and the ones to the left to the $m_R^2=10^{-7}M_{pl}^2$ case. The three $R_0$ cases are shown for each mass: $R_0=0$ (green), $R_0=10^{-6} M_{pl}^2$ (violet) for the smaller mass and  $R_0=10^{-2} M_{pl}^2$ (blue) for the larger one, and  $R_0=4\Lambda_R$ (red).
[Top] The couplings are fixed $\xi_R=0$ and $\lambda_R=0.1$. 
[Middle] We fixed  $\xi_R=0$ and we varied $\lambda_R=0.1$ (solid lines) and $\lambda_R=1$ (dashed lines).
[Bottom] We fixed $\lambda_R=0.1$ and we plotted $\xi_R=0$ (solid lines) and $\xi_R=10^{-6}$ (dashed lines).  The part of the curves where the solid or dashed lines are replaced by grey dotted lines correspond  to   points  for  which the infrared condition $\Delta_m^2\leq\Delta_c^2$ is   violated. On the left panel, there are curves that cannot be distinguished by eye: In all plots, the violet and green curves   are almost superimposed, and in the plot on the bottom,   all  dashed   curves  are also indistinguishable by eye.  
}
        \label{Back3Back5A}
\end{figure}

%\begin{figure}[h!]
%\includegraphics[width=\linewidth]{AAback.pdf}

%\caption{  $\Delta_R=\frac{R-R_{cl}}{R_{cl}}$ vs $\Lambda_R$, for the three $R_0$ cases: $R_0=0$ (green), $R_0=0.01M_{pl}^2$ (blue) y $R_0=4\Lambda_R$ (red). [Top] 
%The couplings are fixed $\xi_R=0$ and $\lambda_R=0.1$ and the masses are  $m_R^2=10^{-7}M_{pl}^2$ (solid lines) and $m_R^2=10^{-3}M_{pl}^2$ (dotted lines).
%[Middle] We fixed  $\xi_R=0$, $m_R^2=10^{-7}M_{pl}^2$ and we varied $\lambda_R=0.1$ (solid lines) and $\lambda_R=1$ (dotted lines).
%[Bottom] We fixed $\lambda_R=0.1$ and $m_R^2=10^{-7}M_{pl}^2$, and we plotted $\xi_R=0$ (solid lines) and $\xi_R=0.01$ (dotted lines).}
%        \label{Back3Back5A}
%\end{figure}
\end{widetext}

Here, we  focus on square masses below $10^{-3} M_{pl}^2$. The reason is because although the IR approximation is expected to be good for  $m_{ph}^2=10^{-3}M_{pl}^2$, one may expect deviations from the approximate solutions if $m_{ph}^2$ is larger. For $m_R^2=10^{-7} M_{pl}^2$, from  Fig. \ref{mph1to4} we see that although $\Delta_{m^2}$ reaches values as large as $10^4$ for $R_0=0$ and $\lambda_R=0.1$, in the plotted range of $R$, the physical mass reaches values at most of order  $10^{-3} M_{pl}^2$. However, for $m_R^2=10^{-3} M_{pl}^2$ (also for  $\lambda_R=0.1$ in the plotted range of $R$), we obtain the physical mass can reach values about $2\times 10^{-3} M_{pl}^2$ for both $R_0=0$ and $R_0=10^{-2} M_{pl}^2$, or remain closer to $10^{-3}M_{pl}^2$ for $R_0=4\Lambda_R$. 

As in Fig.  \ref{mph1to4},  in Fig. \ref{Back3Back5A}  the case  $R_0=4 \Lambda_R$ is to be interpreted with care, since each point on the (red) curves, corresponds to a different $\Lambda_R$ and therefore  to a different $R_0$, that is, to a different definition of the parameters $m_R$, $\lambda_R$, and $\xi_R$. Notice however, this does not alter the conclusions one can deduce for each fixed $\Lambda_R$.

From the analysis above, we conclude  the backreaction of quantum fields in the IR regime is in all cases perturbatively small. We also conclude that  to study the  quantum backreaction effects, the most convenient choice for $R_0$ at the renormalization point is $R_0=R_{cl}$. This choice  allows for a physically meaningful way of defining the parameters of the theory,   providing  a robust   characterization   of  fields in the IR regime (once the renormalized mass of the field is given, in the absence of large quantum   effects, the physical mass remains close to the renormalized one, as expected).
    Using this choice,   in this section  we obtained  the resulting curvature is in all cases smaller than the classical one (and this does not occur for the other studied $R_0$ values). According to the parametrization of the theory in such scheme, this result can be interpreted as the so-called 'screening of the cosmological constant  $\Lambda_R$' by  quantum effects of  IR fields.
%For the same value of $\Lambda_R$, the curvature scalar $R$ value is   reduced due to the quantum effects of the scalar fields. In another words, for the same value of $\Lambda_R$,   the quantum effects screen $R_{cl}$ (ie., the resulting curvature scalar $R$ is smaller than the classical one $R_{cl}=4\Lambda_R$).

\section{Comparison  with previous works}\label{ComparacionConAraiYSerreau}
 
Some of the  results we have obtained can be compared with previous work. In particular, we consider here the results presented in \cite{Arai} and \cite{Serreau}. Both papers   consider the same scalar field theory as we are considering here (given by Eq.~(\ref{ModelSF})   in the semiclassical large $N$ approximation,  and both found a  screening phenomenon of the cosmological constant, but using alternate approaches. 
In \cite{Arai}  the same 2PI EA formalism in the large $N$ limit is considered, while in   \cite{Serreau} the analysis of the backreaction is done using the so-called   Wilsonian  renormalization group framework. The main difference is in their conclusion on the size and parametric dependence of the quantum backreaction effects for light fields.
While  \cite{Arai} concludes  the backreaction is nonpertubatively large, obtaining an unsuppressed effect proportional to a  logarithmic enhancement factor $\log {\lambda}$,  the conclusion in \cite{Serreau} indicates there is no enhancement factor as $\lambda\to 0$ and that the corrections are controlled in the semiclassical approximation by a factor of $R/M_{pl}^2$.

As we have shown above, by computing the renormalized parameters, finding both the physical mass equations and the SEE as a function of these and numerically solving both of these equations we have ultimately found the screening. However, in  \cite{Arai} and \cite{Serreau} the results are shown in terms of the minimal subtraction (MS) parameters.
 In order to compare our result with theirs, we set   $m=0$, $\xi=0$ and write our results (now expressed in terms of the dSR renormalized parameters ($m_R$,  $\xi_R$, $\lambda_R$) and $R_0$), in terms of $\lambda$ and $\hat{\mu}$. 
 It is important to remember a couple of things. The fact that the MS mass $m$ is set to zero is possible here, as long as $m_{ph}$ remains a real and positive. When setting $R=R_0$ (at the renormalization point),  we obtain   $m_R=m_{ph}$, as seen in (\ref{mphR0cte}).
 Therefore, by fixing the curvature $R$ to be equal to the one at the renormalization point $R_0$ (i.e., by setting $R=R_0$) we can obtain the physical mass (also called dynamical mass in the literature) as a function of $R_0$ from the relations given in Eqs.~(\ref{mR2R0cte}), (\ref{xiRR0cte}) and (\ref{lambdaRR0cte}). 
 The computations of $m_R$ and $\xi_R$ in this limit can be seen in Appendix B. In this case, our result for the so-called dynamical mass ($m_{dyn}=m_R=m_{ph}$ for $R=R_0$) can be immediately compared with previous  calculations   performed in the literature in the MS scheme, and one finds it  agrees with them. After inserting them in our expression for  $\langle T_{\mu\nu}'\rangle=\langle T_{\mu\nu}\rangle_{ren}+\langle {T_{\mu\nu}} \rangle_{ad4}^{con}$ and expanding in  $\sqrt{\lambda}$, we obtain 
\begin{equation}
\begin{aligned}
   &\langle T_{\mu\nu}'\rangle\simeq g_{\mu\nu}N\left[\frac{29}{17280\pi}R^2\right.
   -\frac{R^2}{384\sqrt{3}\pi}\sqrt{\lambda}\\
   &+\frac{R^2}{110592\pi^3}\Bigg{\{}95-24\gamma_E
  \left.\left.+12\log\left(\frac{R}{12\hat{\mu}^2}\right) \right\}
   \lambda+...\right],
   \label{TmunuComparacion}
   \end{aligned}
\end{equation}
where we have used that $R_0=R$.

 Therefore,  our result disagree with the one   presented in  \cite{Arai}, on that we do not find a term proportional to $R_0^2\log( \lambda)$, which would generate a large backreaction effect for small values of $\lambda$. As one can see in (\ref{TmunuComparacion}), the contributions involving the coupling constant  are suppressed by  $\sqrt{\lambda}$,  and   do   not result in a large backreaction effect. 
Notice that  Eq.~(\ref{TmunuComparacion}) shows the next to leading order in $\sqrt{\lambda}$ is the one depending on $\hat{\mu}$, but the leading order is independent on  $\hat{\mu}$.  We see that, provided  $\lambda\ll1$, the dominant contribution to the LHS of the SEE [see Eq. ~(\ref{EES_DS_Renormalizadas})] is positive,   leading to the phenomenon of screening  (i.e., $R$ is smaller than $R_{cl}=4\Lambda_R$). All contributions are suppressed by $R/M_{pl}^2$. Therefore, our results agree with the ones presented in \cite{Serreau} and   differ from those in \cite{Arai}.  

As far as we understand the discrepancy is due to a mistake in the procedure followed in \cite{Arai} to compute the SEE from the 2PI EA, after having evaluated the action for a dS metric.   
 Indeed, the   effective action (which in dS is given by the effective potential)  obtained in 
 \cite{Arai} is compatible with ours. The expression for the physical mass (named dynamical mass in \cite{Arai}) also agrees. However, as done in \cite{Serreau}, the correct procedure to obtain  the SEE from the effective action  evaluated in a dS metric, with curvature $R=12 H^2$, is to  perform the derivative $\partial_H(H^{-d}V)=0$, with $d=4$ and  $V=V(H)= (2k_R)^{-1}(12 H^2-2\Lambda_R)+V_{eff}(H)$, where $V_{eff}$ is the ($H-$dependent) scalar field effective potential. In  \cite{Arai}, however, it seems   the factor $H^{-d}$ was not included in the derivation and this gives the extra term with the logarithmic factor.

 As a final remark of the section, it is worth noticing that our results agree with the conclusion of \cite{RevisitedI,RevisitedII} where, as mentioned above, the analysis of renormalization schemes with different values of $R_0$ was done for a single field in the Hartree approximation. In particular, in \cite{RevisitedII} a conclusion was that for sufficiently large values of $R_0$
 the approximation $m_{ph}^2+\xi_R R\ll R$ does not break down as $\Lambda_R\to 0$ (i.e., as $R_{cl}\to 0$), and 
 there is a divergence of the relative deviation $\Delta_R$ in this limit due to the backreaction. In that case it can be seen that as $R_{cl}\to 0$, the curvature $R$ goes to a finite positive value. Therefore, there are parameters for which the backreaction is crucial to determine the spacetime curvature.

\begin{figure} 
\includegraphics[width=\linewidth]{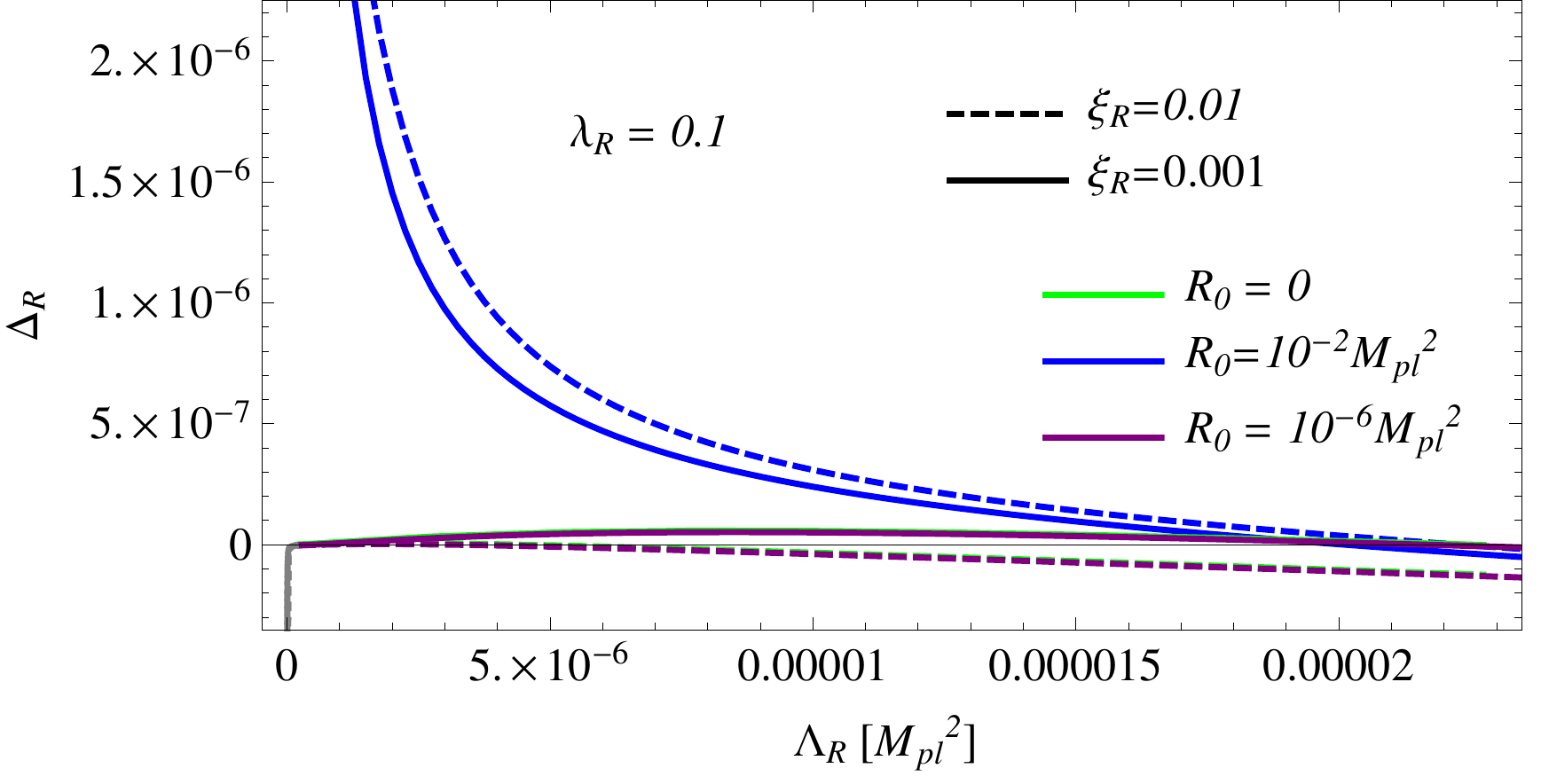}
\caption{ $\Delta_R$ vs  $\Lambda_R$ for $m_R^2=10^{-7}M_{pl}^2$, $\Lambda_R=0.1$, two different values of $\xi_R$ ($\xi_R=0.01$ with dashed lines and $\xi_R=0.001$ with solid lines) and three values of $R_0$: $R_0=0$ (green), $R_0=10^{-6} M_{pl}^2$ (violet) and $R_0=10^{-2}M_{pl}^2$ (blue). The curves for $R=0$ and  $R=10^{-6} M_{pl}$ are almost superimposed.   }        \label{CompLeo}
  
\end{figure}

  Indeed,  as can be seen in Fig. \ref{CompLeo}
   we obtain    solutions for which  $\Delta_R$ diverges at small $\Lambda_R$ for the corresponding parameters ($m_R^2=10^{-7} M_{pl}^2$, $\lambda_R=0.1$ and  for two values of $\xi_R$: $\xi_R=0.01$ and $\xi_R=0.001$\footnote{Recall that when $R_0=10^{-2} M_{pl}^2$, as we mentioned in Sec. \ref{subphysM}, for $m_R^2=10^{-7} M_{pl}^2$ and $\xi_R=0$ the physical mass is not well defined for small values of $R$. For this reason here we only consider cases  with $\xi_R\neq0$.}) for the largest value of $R_0=10^{-2} M_{pl}^2$. For the smaller values of $R_0$ ($R_0=0$ and $R_0=10^{-6}M_{pl}^2$ in the plot) the approximation $m_{ph}^2+\xi_R R\ll R$ breaks down for small values of $R$, as is also seen in Fig. \ref{Back3Back5A}.
   
   We notice however  that the solutions for which $\Delta_R$ diverge only show up in cases where $R_0$ is fixed to be sufficiently large and therefore very different from $R_{cl}=4\Lambda_R$ (as $\Lambda_R\to 0$).  The physical interpretation of the  parametrization of the theory  obtained as $\Lambda_R\to 0$  and the corresponding characterization of the  quantum corrections is unclear to us.  
   Our main focus here has been  to study  the importance of quantum effects in a physically meaningful parametrization of the theory, in order to understand whether or not (or under which conditions) the quantum backreaction (although small in size)  contribute to  screen the classical solution.  

\section{Conclusions}
    
   The main subject of  the present work has been the  problem of the    backreaction  of quantum fields  on the spacetime  curvature through the SEE. We focus on  the $O(N)$ theory defined in Eq.~(\ref{ModelSF})  in the symmetric phase of the field (i.e., for vanishing vacuum expectation values of the scalar fields), in the large $N$ approximation.

A main result of this paper is a set of finite (renormalized) self-consistent equations for backreaction studies in a general background metric in the de Sitter   renormalization (dSR) schemes defined in Sec.~\ref{dSR}, namely: the renormalized  gap-equation in Eq.~(\ref{mphR0cte}), necessary to obtain finite equations for the  fields  and the two point functions (i.e., Eqs.~(\ref{EquMovPhi}) and  (\ref{EquMovG})), and the renormalized  SEE presented in Sec.~\ref{SecRenSEE}. For $N=1$, these equations reduce to the ones obtained in Ref. \cite{RevisitedII} in the Hartree approximation, in the symmetric phase, up to a rescaling of the coupling constant by a factor $3$ (i.e., $\lambda_R\to 3 \lambda_R$). We emphasize the fact that these equations can be used as a starting point to study the quantum backreaction problem beyond dS spacetimes, since dS is used only as an alternative  spacetime at the renormalization point. This choice generalizes the traditional one that uses Minkowski geometry at the renormalization point. As happens for the light fields in dS spacetime considered in this paper, this generalization could be significantly useful when infrared effects are sensitive to the curvature of the spacetime. We point out the use of dSR  schemes may be  also useful for non-dS geometries, such as for more generic Friedman Robertson Walker spacetimes used in cosmology. 

Another important result is the specific study of the quantum backreaction problem for dS spacetimes. This  has allowed us to explicitly illustrate the importance of the dSR schemes in the understanding of the physical results. More specifically,   we have obtained  a system of two equations  (Eq. (\ref{mphR0cte}) and Eq.(\ref{EES_DS_Renormalizadas})) that can be  solved numerically  to assess the effects on the curvature due to the presence of quantum fields for different values of the renormalized    parameters of the fields (i.e., the mass $m_R$, the coupling constant $\lambda_R$ and  the coupling to the curvature $\xi_R$) and the renormalized cosmological constant $\Lambda_R$. 

First, we have analyzed the impact of choosing the geometry at the renormalization point in the relative difference between the physical mass and the renormalized mass as the physical background geometry (characterized by the curvature scalar $R$ in this case)  changes. We have obtained the difference is minimal, for a wide range of values of $R$ in the semiclassical approximation, when the dS geometry fixed  at the  renormalization point is  the classical solution, $R_0=R_{cl}=4\Lambda_R$.
This indicates the definition of the physical mass is less sensitive to curvature variations and changes in the  interaction between the quantum fields and their coupling to the curvature. Hence, we conclude  the  choice of this dSR scheme (with $R_0=R_{cl}$) is more convenient to study the quantum  backreaction effects than choosing the plane geometry $R_0=0$ or   another fixed value for $R_0$. 

Then, we have studied  the relative difference between the curvature (that is affected by the quantum interactions between the fields) and its classical approximation (where the quantum effects are neglected). We have found that for light fields this difference is always negative, that is, that the curvature is  smaller than the classical one. For the other  mass values analyzed, this phenomenon has  also been found, but only for the dSR scheme with $R_0=R_{cl}$ case.  Given the previous conclusion on this dSR scheme regarding the sensitivity to quantum physics, we consider this result can then be interpreted as an screening of the cosmological constant induced by quantum effects.

 \acknowledgments
 We thank D.~F.~ Mazzitelli and L.~G.~ Trombetta for useful comments and discussions. 
 DLN has been supported by CONICET, ANPCyT and UBA.

\appendix

\section{Renormalization counterterms}

In order  to obtain a nondivergent equation for the physical mass, we absorb the divergencies into  the bare parameters of the theory ($m_B$, $\xi_B$ and $\lambda_B$). We have

\begin{equation}
    m_{ph}^2+\xi_RR=m^2+\delta m^2+(\xi+\delta\xi)R+\frac{1}{4}(\lambda+\delta\lambda)[G_1].
\end{equation}

Therefore, using    Eq. (\ref{G1coincidence}) for  $[G_1]$, 
\begin{equation}
    \begin{aligned}
    m_{ph}^2+\xi_RR&=m^2+\delta m^2+(\xi+\delta\xi)R\\
    &+\frac{1}{16\pi^2\epsilon}(\lambda+\delta\lambda)\left[m_{ph}^2+\left(\xi_R-\frac{1}{6}\right)R\right]\\
     & +\frac{1}{2}(\lambda+\delta\lambda)T_F
    \end{aligned}
\end{equation} and demanding that the divergent terms cancel out, we obtain

\begin{equation}
    \begin{aligned}
    0=&\left[ \delta m^2+\frac{m^2}{16\pi^2\epsilon}(\lambda+\delta\lambda) \right]+\frac{1}{2}\left[\frac{\lambda}{16\pi^2\epsilon}(\lambda+\delta\lambda) +\right.\\
    &\delta\lambda \Bigg{]}T_F
      +\left[\delta\xi+\frac{1}{16\pi^2\epsilon}\left( \xi-\frac{1}{6} \right)(\lambda+\delta\lambda) \right]R.
    \end{aligned}
\end{equation}
Therefore, the required counterterms are:
\begin{equation}
    \delta m^2=-\frac{\lambda}{16\pi^2\epsilon}\left(\frac{m^2}{1+\frac{\lambda}{16\pi^2\epsilon}}\right),
\end{equation}

\begin{equation}
    \delta \xi=-\frac{\lambda}{16\pi^2\epsilon}\left(\frac{(\xi-\frac{1}{6})}{1+\frac{\lambda}{16\pi^2\epsilon}}\right),
\end{equation}

\begin{equation}
    \delta \lambda=-\frac{\lambda}{16\pi^2\epsilon}\left(\frac{\lambda}{1+\frac{\lambda}{16\pi^2\epsilon}}\right).
\end{equation}

\section{$\langle T_{\mu\nu} \rangle$ at LO in $\sqrt{\lambda}$}

In Refs.~\cite{Arai} and \cite{Serreau} the results are presented in terms of the MS parameters $(m,\xi,\lambda)$ and $\hat{\mu}$. However, we present our  results   in terms of renormalized parameters  $(m_R,\xi_R,\lambda_R)$ in the dSR scheme with a generic $R_0$. In order to compare the results, in this appendix we set $m=\xi=0$  and $R=R_0$, and
provide a relation  between $(m_R,\xi_R,\lambda_R)$  and $\lambda$ and $\hat{\mu}$, valid at the next to leading order in an expansion in $\sqrt{\lambda}$. Recall that, by definition and  since we are setting $\hat{\phi}=0$,  the renormalized mass $m_R$ is the physical mass $m_{ph}$ at $R=R_0$. From Eq. (\ref{mR2R0cte}), using Eq. \eqref{FdS} and Eq. \eqref{gLimIR} we have \cite{RevisitedI}

\begin{equation}
\begin{aligned}
    m_R^2&= -R_0\left(\xi_R-\frac{1}{6}\right)\\
    &-\frac{\frac{R_0}{6}+\frac{\lambda R_0}{576\pi^2}}{1-\frac{\lambda}{32\pi^2}\left[ \log\left(\frac{R_0}{12\hat{\mu}^2}\right)+g\left(\frac{m_R^2+\xi_RR_0}{R_0}\right)\right]},
    \end{aligned}
\end{equation}

Then,   at NLO in $\sqrt{\lambda}$, it is reduced to a quadratic equation on $m_R^2$, whose solution is

\begin{equation}
    m_R^2\simeq\frac{R_0\sqrt{\lambda}}{16\sqrt{3}\pi}+\lambda R_0\left[\frac{12\gamma_E-13+6\log\left(\frac{R_0}{12\hat{\mu}^2}\right)}{2304\pi^2}\right].
\end{equation}

Performing the analogous procedure for $\xi_R$, we obtained 

\begin{equation}
    \xi_R\simeq\frac{\sqrt{\lambda}}{16\sqrt{3}\pi}+\lambda\left[\frac{6\gamma_E-14-3\log\left(\frac{R_0}{12\hat{\mu}^2}\right)}{1152\pi^2}\right].
\end{equation}

Hence, when inserting these results for $m_R$ and $\xi_R$ into Eq.~(\ref{RHS_EES_DS}), and considering the NLO terms in $\sqrt{\lambda}$ expansion, we get the result shown in Eq. (\ref{TmunuComparacion}).

\bibliography{Biblio}

\end{document}